\documentclass[aps,pra,reprint,superscriptaddress,english]{revtex4-1}

\usepackage{color}
\usepackage{esint}
\usepackage{graphicx}% Include figure files
\usepackage{siunitx} %SI Units
\usepackage{braket} % for bras and kets of quantum states
\usepackage{hyperref}
\hypersetup{
    colorlinks=true,
    linkcolor=black,
    citecolor=black,
    filecolor=black,
    urlcolor=black,
}

\begin{document}

\title{Converting one photon into two via four-wave mixing in optical
fibers}
\author{Audrey Dot}
\email{audreyddot@gmail.com}
\affiliation{Institute for Quantum Computing and Department of Physics and Astronomy, University of Waterloo, 200 University Ave W, Waterloo, Ontario, Canada, N2L 3G1}
\author{Evan Meyer-Scott}
\affiliation{Institute for Quantum Computing and Department of Physics and Astronomy, University of Waterloo, 200 University Ave W, Waterloo, Ontario, Canada, N2L 3G1}
\author{Raja Ahmad}
\affiliation{Department of Electrical and Computer Engineering, McGill University, 3480 rue University, Montreal, Quebec, Canada, H3A 2A7}
\author{Martin Rochette}
\affiliation{Department of Electrical and Computer Engineering, McGill University, 3480 rue University, Montreal, Quebec, Canada, H3A 2A7}
\author{Thomas Jennewein}
\affiliation{Institute for Quantum Computing and Department of Physics and Astronomy, University of Waterloo, 200 University Ave W, Waterloo, Ontario, Canada, N2L 3G1}

\begin{abstract} Observing nonlinear optical quantum effects or implementing quantum information protocols using nonlinear optics requires moving to ever-smaller input light intensities. However, low light intensities generally mean weak optical nonlinearities, inadequate for many applications. Here we calculate the performance of four-wave mixing in various optical fibers for the case where one of the input beams is a single photon. We show that in tapered chalcogenide glass fibers (microwires) a single photon plus strong pump beam can produce a pair of photons with probability 0.1\%, much higher than in previous work on bulk and waveguided crystal sources. Such a photon converter could be useful for creating large entangled photon states, for performing a loophole-free test of Bell's inequalities, and for quantum communication.\end{abstract}

\maketitle

\section{Introduction}
Pairs of photons created via Spontaneous Parametric Down-Conversion (SPDC)~\cite{Burnham_SPDC_70} or spontaneous Four-Wave Mixing (FWM)~\cite{Fiorentino2002-sp-FWM} in a nonlinear optical material with a high-intensity pump laser have been used in many experiments in quantum optics, quantum metrology, and optical quantum information processing. Interest is increasingly converging on using SPDC or FWM in later stages of quantum information protocols, rather than just initial sources of photons~\cite{Guerreiro2013Interact,2014arXiv1404.7131H,PhysRevLett.113.013601}. This requires operation with very low intensity input states, including converting a single photon into a pair.

Should an efficient one-to-two photon conversion be realized, one important application is the entangling of three or more  photons~\cite{Hubel2010Direct-g,Shalm2013Three-ph}. These large entangled photon states are useful in quantum communication protocols~\cite{Browne_Rudolph2005,Hillery_1999},
and allow fundamental tests of quantum mechanics~\cite{banaszek1997qit,greenberger1990bst,Shalm2013Three-ph}. Increased efficiency in converting single photons to pairs would allow larger states to be generated, and with greater speed. Single photon conversion could also be used for heralding photons after long-distance transmission to close the Bell test detection loophole~\cite{PhysRevX.2.021010} and for device-independent quantum key distribution~\cite{PhysRevLett.105.070501}; any improvement in conversion efficiency directly increases the communication rates. Finally, if efficient enough, single photon conversion could also be used directly in quantum computing as a two-qubit gate~\cite{Langford2011Efficien}.

The key challenge in converting a single photon into a pair is the low efficiency of nonlinear optical processes at ultralow power. In principle, standard SPDC or FWM sources could be used, but the low efficiency (less than $10^{-5}$) limits the single photon conversion to rates too low to be useful~\cite{Tanzilli4Highly-e}. Therefore we consider here specialty fiber media, which we show can result in large conversion efficiencies thanks to long length, small core size, and high nonlinearity.

We present complete calculations and simulations of FWM between a strong
pump and a single photon as illustrated schematically in Fig.~\ref{fig:scheme}. First we set up the theoretical framework by extending the quantum theory of nondegenerate FWM~\cite{Brainis2009,Agrawal2007} 
to the single photon pump case.
Then we apply the expressions to birefringent silica fibers, microstructured silica fibers, and chalcogenide microwires to find the spectra and conversion efficiency of the generated two-photon states. 

\section{Quantum theory of FWM pumped by a single photon and strong laser}

\subsection{Equations of motion for a $\chi^{(3)}$ nonlinear medium}

\begin{figure}
\includegraphics[width = \columnwidth]{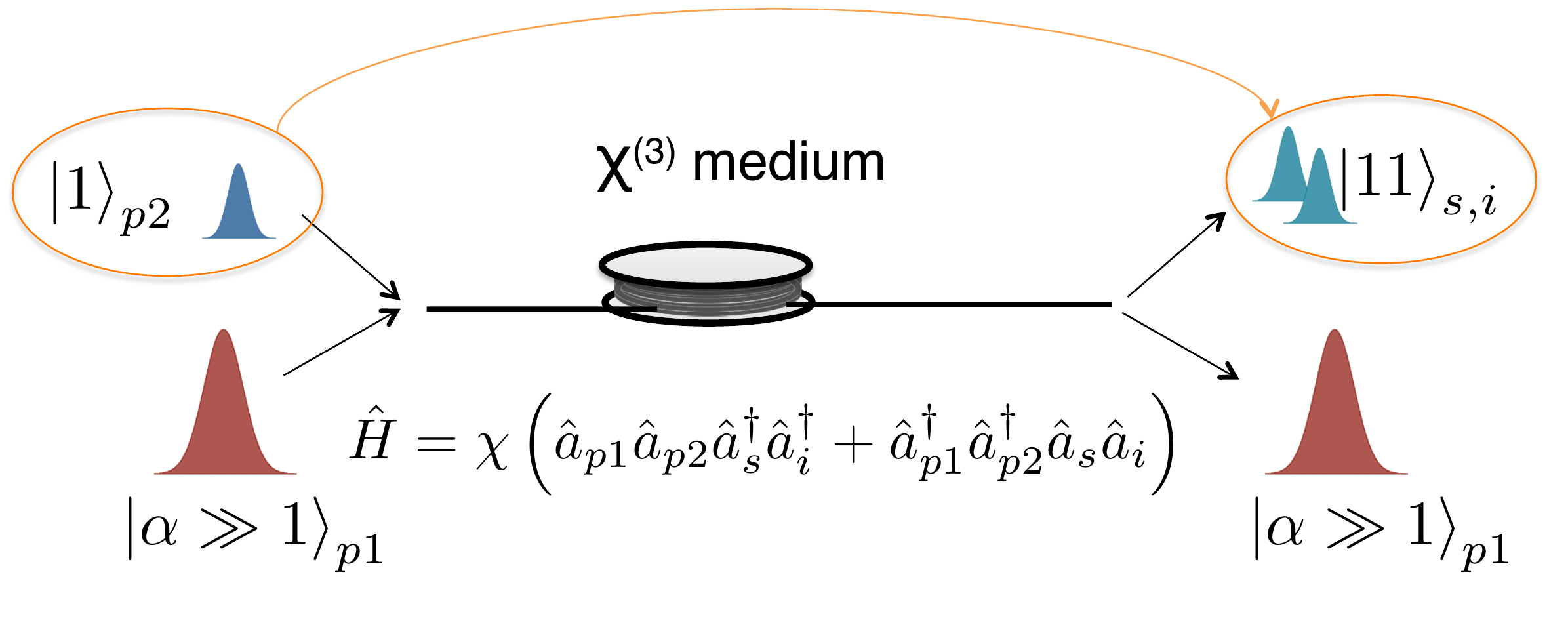}
\caption{[Color online] Four-wave mixing (FWM) with single photon and strong pump inputs. The toy Hamiltonian with interaction parameter $\chi$ illustrates the processes, but the full development is given in the text.\label{fig:scheme}}
\end{figure}

We determine the operator evolution of a system consisting of a strong pump beam
and a single photon entering a nonlinear, dispersive, single mode fiber,
and undergoing four-wave mixing and phase modulation as in Fig.~\ref{fig:scheme}. We proceed in the
Heisenberg representation by solving the equation of motion for the
field amplitude operators~\cite{Huttner1990}. The two pumping fields are considered monochromatic or quasi-monochromatic, 
and we include self- and cross-phase modulation, but not parasitic effects such as Raman scattering and multi-photon absorption due to the low power of the inputs. We stay in the low gain
regime, which means that only spontaneous FWM (also called Four Photon Scattering)
is studied. This approximation stands if the total probability of generating a photon pair during the interaction is much lower than 1, and certainly holds since one of our pumps is a single photon.

The field is quantized in one dimension in a length large enough for the electric field to be written in the continuous limit~\cite{Loudon1990}. We then choose for convenience to write this field in the frequency space, as a sum of its space-dependent spatial mode operators, an approach introduced in Ref.~\cite{Huttner1990}. The quantization time $T$, equal to the quantization length divided by the speed of light, is then the time periodicity of the field, and the density of the frequency space is $\delta\omega=2\pi/T$. $T$ has to be long enough to allow the writing of the frequency modes in the continuous limit. The electric field is then
\begin{eqnarray}
\hat{\vec{E}}(r,t)  =  &\sum_{j=x,y}&\left(F(x,y)\sqrt{\frac{\text{\ensuremath{\hbar}}}{2\varepsilon_{0}c}}\frac{1}{\sqrt{2\pi}}\right.\\
&  &\times\left.\int d\omega\sqrt{\frac{\omega}{n_{j}(\omega)}}\hat{a}_{j}(\omega,z)e^{-i\ensuremath{\omega}t}+h.c.\right)\vec{e_{j}}\label{eq:field1},\nonumber
\end{eqnarray}
where the frequency integral runs from 0 to $+ \infty$, and "$h.c.$" stands for hermitian conjugate. $F(x,y)$, with $\iint\left|F(x,y)\right|^{2}dxdy=1$,
is the transverse distribution of the fiber mode. The unit vectors $\vec{e_{j}}$ describe the field's polarization, and $n_{j}(\omega)$ is the effective index of refraction for the fiber mode of frequency $\omega$ and polarization $j$. Since in the continuous limit, we approximated the discrete longitudinal modes of a laser cavity by continuous mode annihilation operators, $\hat{a}_{j}(\omega,z)$, with units of inverse square root frequency~\cite{Loudon1990}. They follow the commutation
relations $\left[\hat{a}_{j}(\omega,z),\hat{a}_{j'}^\dagger(\omega',z)\right]=\delta(\omega-\omega')\delta_{jj'}$, with Dirac delta $\delta(\omega-\omega')$ and Kroenecker delta $\delta_{jj'}$.

These operators $\hat{a}_{j}(\omega,z)$
and the quantum state of the system $\left|\psi\right\rangle$ provide
complete knowledge about the state as a function of propagation distance $z$, which allows us to extract the efficiency of single photon to pair conversion. In the Heisenberg representation, $\ket{\psi}$ is constant and we only need to solve for the evolution of the annihilation operators.

If considering only one polarization component, the propagating field can be simplified into
\begin{eqnarray}
\hat{E}(z,t) & = & \sqrt{\frac{\text{\ensuremath{\hbar}}}{4\pi\varepsilon_{0}cA_{\text{eff}}}}\int d\omega\sqrt{\frac{\omega}{n(\omega)}}\hat{a}(\omega,z)e^{-i\ensuremath{\omega}t}+h.c.\label{eq:Esimpl}\nonumber\\
 & = & \hat{E}^{(+)}(z,t)+ \hat{E}^{(-)}(z,t),
\end{eqnarray}
where the transverse modal distribution was also simplified using the effective area of
the fiber mode $A_{\text{eff}}=\frac{1}{\int\int|F(x,y)|^4 dxdy}$, taken
to be the same for all the frequency components in the fiber.

The evolution equation of the annihilation operators can be given by 

\begin{equation}
\frac{\partial\hat{a}(\omega,z)}{\partial z}=\frac{i}{\hbar}\left[\hat{a}(\omega,z),\hat{G}(z)\right]\label{eq:da/dz},
\end{equation}
where the momentum operator $\hat{G}$ is given by integration over the cross-sectional area of the momentum that flows during the quantization time.

\begin{eqnarray}
\hat{G}(z) & = & \int_{A_{\text{eff}}}dS\int_{0}^{T}dt\hat{D}^{(-)}(z,t)\hat{E}^{(+)}(z,t)+h.c., \label{eq:g}
\end{eqnarray}
where $\hat{D}(z,t)=\varepsilon_{0}\hat{E}(z,t)+\hat{P}(z,t)$ is
the electric displacement operator. The polarization operator is defined by

\begin{eqnarray}
\hat{P}(z,t) & = & \sum_{n\geq1}\varepsilon_{0}\chi^{(n)}\cdot\hat{E}^{n}(z,t)\label{eq:P}\\
 & = & \hat{P}_{l}(z,t)+\hat{P}_{nl}(z,t)\nonumber,
\end{eqnarray}
which is the sum of the linear polarization $\hat{P}_{l}(z,t)$ given by
$\varepsilon_{0}\chi^{(1)}(\omega)\cdot\hat{E}(r,t)$, and the nonlinear
polarization $\hat{P}_{nl}(z,t)$ of higher orders. We can thus
separate $\hat{G}(z)$ into linear and nonlinear parts as $\hat{G}(z)=\hat{G}_{l}(z)+\hat{G}_{nl}(z)$, driven by corresponding linear and nonlinear polarizations. 

The linear evolution of the momentum operator is obtained from Eqs.~(\ref{eq:Esimpl}), (\ref{eq:g}),
and (\ref{eq:P}) as (see Appendix~\ref{app:momentum})
\begin{equation}
\hat{G}_{l}(z)=\int d\omega\hbar\beta(\omega)\hat{a}^{\dagger}(\omega,z)\hat{a}(\omega,z),\label{Gl}
\end{equation}
with the propagation constant $\beta(\omega)=\frac{n(\omega)\omega}{c}$.  The linear evolution of any annihilation operator can thus be deduced
from Eqs.~(\ref{eq:da/dz}) and (\ref{Gl}) as
\begin{equation}
\hat{a}_{l}(\omega,z)=\hat{a}_{0}(\omega,z)e^{i\beta(\omega)z}\label{eq:a_lin}.
\end{equation}

The nonlinear evolution (contained in $\hat{a}_{0}(\omega,z))$
can be found similarly from the nonlinear evolution of the momentum.
$\hat{G}_{nl}$ can be decomposed into two parts as $\hat{G}_{nl}(z)=\hat{G}_{nl}^{FWM}(z)+\hat{G}_{nl}^{ph\, mod}(z)$ (see Appendix~\ref{app:momentum}), one giving FWM and the other phase modulation. 

The two pumping fields, of frequencies  $\omega_{p1,2}$, are considered monochromatic or quasi-monochromatic,
perfectly overlapping in time, and have the same spectral bandwidth $\delta\omega_{p}$ with $\delta\omega_{p}/\omega_{p1,2}\ll1$. It is convenient to choose the quantization time as the Fourier transform of the pulses' spectral width,  $T=2\pi/\delta\omega_{p}$. The frequency-space density is therefore $\delta\omega=\delta\omega_p$. For monochromatic pumps, this quantization
time, as well as the pulse duration, is infinite.

The FWM part of the momentum operator is (see Appendix~\ref{app:momentum})
\begin{widetext}
\begin{eqnarray}
 &  & \hat{G}_{nl}^{FWM}(z)=3\chi^{(3)}\frac{\text{\ensuremath{\hbar^2}}}{\varepsilon_{0}c^{2}A_{\text{eff}}T}\times\frac{2\pi}{T}\label{eq:GFWM}\nonumber\\
 &  & \left[\int d\omega\sqrt{\frac{\omega\omega_{p1}\omega_{p2}\left(\omega_{p1}+\omega_{p2}-\omega\right)}{n(\omega)n(\omega_{p1})n(\omega_{p2})n(\omega_{p1}+\omega_{p2}-\omega)}}\hat{a}_{0}^{\dagger}(\omega,z)\hat{a}_{0}^{\dagger}(\omega_{p1}+\omega_{p2}-\omega,z)\hat{a}_{0}(\omega_{p1},z)\hat{a}_{0}(\omega_{p2},z)e^{-i\Delta kz}+h.c.\right]
\end{eqnarray}
\end{widetext}
with $\Delta k=\beta(\omega)+\beta(\omega_{p1}-\omega_{p2}-\omega)-\beta(\omega_{p1})-\beta(\omega_{p2})$, and where the integral over $\omega$ covers the whole positive spectrum
except the two injected frequencies. The two creation operators $\hat{a}_{0}^{\dagger}(\omega,z)\hat{a}_{0}^{\dagger}(\omega_{p1}+\omega_{p2}-\omega,z)$ indicate that output photons can only be created in pairs, with correlated frequencies $\omega$ and $\omega_{p1}+\omega_{p2}-\omega$.

The phase modulation part is (see Appendix~\ref{app:momentum})
\begin{eqnarray}
\hat{G}_{nl}^{ph\, mod}(z) & = & 3\chi^{(3)}\frac{\text{\ensuremath{\hbar}}^{2}}{\varepsilon_{0}c^{2}A_{\text{eff}}T}\left[\iint d\omega d\omega'\frac{\omega}{n(\omega)}\frac{\omega'}{n(\omega')}\right.\nonumber\\
& &\times \hat{a}_{0}^\dagger(\omega,z)\hat{a}_{0}(\omega',z)\hat{a}_{0}^\dagger(\omega',z)\hat{a}_{0}(\omega,z)\label{eq:G_ph_mod} \\
 &  & \left.-\frac{1}{2}\int d\omega\times\frac{2\pi}{T}\left(\frac{\omega}{n(\omega)}\hat{a}_{0}^{\dagger}(\omega,z)\hat{a}_{0}(\omega,z)\right)^{2}\right]\nonumber,
\end{eqnarray}
with the integrals covering the whole positive spectrum.

We can now derive the evolution of the mode operators from Eq.~(\ref{eq:da/dz}),
for any frequency generated in the fiber.
\\

\begin{eqnarray}
 &  & \frac{\partial\hat{a}_{0}(\omega,z)}{\partial z}=3i\chi^{(3)}\frac{\text{\ensuremath{\hbar}}}{\varepsilon_{0}c^{2}A_{\text{eff}}T}\times\label{eq:da/dz_continue}\\
 &  & \left[\frac{2\pi}{T}\sqrt{\frac{\omega\omega_{p1}\omega_{p2}\left(\omega_{p1}+\omega_{p2}-\omega\right)}{n(\omega)n(\omega_{p1})n(\omega_{p2})n(\omega_{p1}+\omega_{p2}-\omega)}}\right.\nonumber\\
 & &\times\hat{a}_{0}^{\dagger}(\omega_{p1}+\omega_{p2}-\omega,z)\hat{a}_{0}(\omega_{p1},z)\hat{a}_{0}(\omega_{p2},z)e^{-i\Delta kz}\nonumber \\
 &  & +\frac{\omega}{n(\omega)}\int d\omega'\frac{\omega'}{n(\omega')}\left[ \hat{a}_{0}^{\dagger}(\omega',z)\hat{a}_{0}(\omega',z)+\frac{1}{2} \frac{T}{2\pi} \right] \hat{a}_{0}(\omega,z)\nonumber\\
 & &-\left.\frac{1}{2}\frac{2\pi}{T}\frac{\omega^2}{n(\omega)^2}\hat{a}_{0}^{\dagger}(\omega,z)\hat{a}_{0}(\omega,z)\hat{a}_{0}(\omega,z)\right]\nonumber.
\end{eqnarray}

The first of the two summed terms reflects the evolution by FWM and the second the self-phase
modulation. If we neglect the phase modulation arising from the generated frequencies as these will be much weaker than the pumps, we have, for the generated frequencies,
\begin{widetext}
\begin{eqnarray}
 &  & \frac{\partial\hat{a}_{0}(\omega,z)}{\partial z}=3i\chi^{(3)}\frac{\text{\ensuremath{\hbar}}}{\varepsilon_{0}c^{2}A_{\text{eff}}T}\times\frac{2\pi}{T}\label{eq:das/dz}\\
 &  & \left[\sqrt{\frac{\omega\omega_{p1}\omega_{p2}\left(\omega_{p1}+\omega_{p2}-\omega\right)}{n(\omega)n(\omega_{p1})n(\omega_{p2})n(\omega_{p1}+\omega_{p2}-\omega)}}\hat{a}_{0}^{\dagger}(\omega_{p1}+\omega_{p2}-\omega,z)\hat{a}_{0}(\omega_{p1},z)\hat{a}_{0}(\omega_{p2},z)e^{-i\Delta kz}+\right.\nonumber \\
  &  & \left.\frac{\omega}{n(\omega)}\left[\frac{\omega_{p1}}{n(\omega_{p1})}\hat{a}_{0}^\dagger(\omega_{p1},z)\hat{a}_{0}(\omega_{p1},z)+\frac{\omega_{p2}}{n(\omega_{p2})}\hat{a}_{0}^\dagger(\omega_{p2},z)\hat{a}_{0}(\omega_{p2},z)+\frac{1}{2}\frac{\omega}{n(\omega)}\frac{T}{2\pi}\right]\hat{a}_{0}(\omega,z)\right]\nonumber .
\end{eqnarray}

For the incoming pump frequencies the evolution is given by

\begin{eqnarray}
 &  & \frac{\partial\hat{a}_{0}(\omega_{j},z)}{\partial z}=3i\chi^{(3)}\frac{\text{\ensuremath{\hbar}}}{\varepsilon_{0}c^{2}A_{\text{eff}}T}\times\label{dap/dz}\\
 &  & \left[ \int d\omega\sqrt{\frac{\omega\omega_{j}\omega_{k}\left(\omega_{j}+\omega_{k}-\omega\right)}{n(\omega)n(\omega_{j})n(\omega_{k})n(\omega_{j}+\omega_{k}-\omega)}}\hat{a}_{0}(\omega,z)\hat{a}_{0}(\omega_{p1}+\omega_{p2}-\omega,z)\hat{a}_{0}^{\dagger}(\omega_{k},z)e^{i\Delta kz}+\right. \nonumber\\
 &  & \left.\frac{2\pi}{T}\frac{\omega_{j}}{n(\omega_{j})}\left[\frac{1}{2}\frac{\omega_{j}}{n(\omega_{j})}\hat{a}_{0}^{\dagger}(\omega_{j},z)\hat{a}_{0}(\omega_{j},z)+\frac{\omega_{k}}{n(\omega_{k})}\hat{a}_{0}^{\dagger}(\omega_{k},z)\hat{a}_{0}(\omega_{k},z)+\frac{1}{2}\frac{\omega}{n(\omega)}\frac{T}{2\pi}\right]\hat{a}_{0}(\omega_{j},z)\right] \nonumber,
\end{eqnarray}
with $j,k = p1,p2$, 
\end{widetext}

Though we are in the quasi-monochromatic approximation, the pumps' creation and annihilation operators are not dimensionless, for homogeneity with those of the generated modes.

\subsection{Solution for a single photon and a strong pump}

To solve Eqs.~(\ref{eq:das/dz}) and (\ref{dap/dz}), the strong pump is taken as classical ($\hat{a}_{0}(\omega_{p1},z)\equiv A_{p1}(z)$),
and undepleted ($\left|A_{p1}(z)\right|^{2}=\left|A_{p1}(0)\right|^{2}$). The weak pump, $p2$, has to be kept quantum throughout, since it is on the few- or single-photon level. Therefore we can also assume the number of weak pump photons is negligible compared to the number of strong pump photons and so neglect phase modulation from the weak pump.

We use the standard waveguide nonlinear parameter $\gamma(\omega)=\frac{3\chi^{(3)}\omega}{2\varepsilon_{0}c^{2}n(\omega)^2 A_{\text{eff}}}$.
If all the frequencies are close, we can use the same, averaged $\gamma$ for
all the frequency modes, which is commonly used to simplify the notation but is not necessary for the solution~\cite{Agrawal-NL-Fibre-Optics}.

With these approximations the evolution of the strong pump can be simplified from Eq.~(\ref{dap/dz})
to 
\begin{equation}
  \frac{dA_{p1}(z)}{dz}= i\gamma P_1 A_{p1}(z)\label{eq:daps/dz},
\end{equation}
where we defined the pump peak power as
$P_{1}(z)=\frac{\hbar\omega_{p1}\times N_{1}(z)}{T}=\frac{2\pi\hbar\omega_{p1}}{T^{2}}\times\left|A_{p1}(z)\right|^{2}$. Here 
$N_{1}(z)=\frac{2\pi}{T}\left|A_{p1}(z)\right|^{2}$
is the number of pump photons going through a plane at position $z$ per time $T$. In the undepleted pump approximation, $P_{1}$ is independent of $z$.

Equation (\ref{eq:daps/dz}) is solved as \footnote{Material absorption can be included here and in all the following evolutions by taking $z$ as the effective position in the fiber; $z=\frac{1-e^{-\alpha z^\prime}}{\alpha}$ for absorption coefficient $\alpha$ and true position $z^\prime$. }
\begin{equation}
A_{p1}(z)=A_{p1}(0)e^{i\gamma P_{1}z}.
\end{equation}

The evolution of the weak pump can then be simplified from Eq.~(\ref{dap/dz}) to

\begin{widetext}
\begin{equation}
\frac{\partial\hat{a}_{0}(\omega_{p2},z)}{\partial z}= \frac{2i\gamma\hbar}{T}\int d\omega\sqrt{\omega\left(\omega_{p1}+\omega_{p2}-\omega\right)}A_{p1}^{*}(0)e^{-i\gamma P_{1}z}\hat{a}_{0}(\omega,z)\hat{a}_{0}(\omega_{p1}+\omega_{p2}-\omega,z)e^{i\Delta kz}+ 2i\gamma P_{1}\hat{a}_{0}(\omega_{p2},z),
\end{equation}
and if we choose $A_{p1}^{*}(0)=A_{p1}(0)= T\sqrt{\frac{P_{1}}{2\pi\hbar\omega_{p1}}}$ then 
\begin{equation}
 \frac{\partial \hat{a}_{0}(\omega_{p2},z)}{\partial z}=2i\gamma\left[ \sqrt{P_{1}}\sqrt{\frac{\hbar}{2\pi\omega_{p1}}}\int d\omega\sqrt{\omega\left(\omega_{p1}+\omega_{p2}-\omega\right)}\hat{a}_{0}(\omega,z)\hat{a}_{0}(\omega_{p1}+\omega_{p2}-\omega,z)e^{i\left(\Delta k-\gamma P_{1}\right)z}+P_{1}\hat{a}_{0}(\omega_{p2},z)\right].
\end{equation}
We can write this more explicitly by introducing $\zeta_{2}=\frac{2\pi\hbar\omega_{p2}}{T^{2}}$, where
$\zeta_{2}\times\left\langle \hat{a}^\dagger(\omega_{p2},0)\hat{a}(\omega_{p2},0)\right\rangle = P_{2}$ is
the peak power of the weak pump at the medium entrance, with $P_2=\zeta_{2}\times\frac{T}{2\pi}$
in case of a single photon pumping. We then have 
\begin{equation}
 \frac{\partial \hat{a}_{0}(\omega_{p2},z)}{\partial z}=2i\gamma\left[ \sqrt{P_{1}}\sqrt{\zeta_{2}} \frac{2\pi}{T}\int d\omega\hat{a}_{0}(\omega,z)\hat{a}_{0}(\omega_{p1}+\omega_{p2}-\omega,z)e^{i\left(\Delta k-\gamma P_{1}\right)z}+P_{1}\hat{a}_{0}(\omega_{p2},z)\right]\label{eq:dap2/dz}.
\end{equation}

Finally, the evolution of the generated modes' annihilation operators, simplified from Eq.~(\ref{eq:das/dz}), is 
\begin{equation}
  \frac{\partial \hat{a}_{0}(\omega,z)}{\partial z}=2i\gamma\left[ \sqrt{\zeta_{2}}\sqrt{P_{1}}\hat{a}_{0}^\dagger(\omega_{p1}+\omega_{p2}-\omega,z)\hat{a}_{0}(\omega_{p2},z)e^{-i\left(\Delta k-\gamma P_{1}\right)z}+P_{1}\hat{a}_{0}(\omega,z)\right].
\end{equation}

The evolution of both the weak pump and of the generated photons can be
derived in the low gain approximation by using a Baker-Hausdorff
expansion to first order in the effective gain $\gamma\sqrt{T\zeta_{2}}\sqrt{P_{1}}L\ll1$. 
The calculations for the annihilation operators of the generated frequencies are detailed in Appendix~\ref{app:BCH} and give the main result
\begin{eqnarray}
& &\hat{a}_{0}(\omega,L)e^{-i2\gamma P_{1}L} =  \hat{a}_{0}(\omega,0)+2i\gamma\times\sqrt{P_{1}}\sqrt{\zeta_{2}}Le^{-\frac{iKL}{2}} \text{sinc}\left(\frac{KL}{2}\right)\times\hat{a}_{0}^{\dagger}(\omega_{p1}+\omega_{p2}-\omega,0)\hat{a}_{0}(\omega_{p2},0)\label{as_solved},
\end{eqnarray}
where $K=\Delta k+\gamma P_{1}=\beta(\omega)+\beta(\omega_{p1}+\omega_{p2}-\omega)-\beta(\omega_{p1})-\beta(\omega_{p2})+\gamma P_{1}$ is the total phase mismatch, sum of the linear and
nonlinear parts.
\end{widetext}

Note that by considering only the first order gain, we assume that the conversion efficiency is low enough to be well represented by a single conversion process, described by  $\hat{a}_0(\omega_{p1},L)\hat{a}_0(\omega_{p2},L)\hat{a}_{0}^{\dagger}(\omega,L) \hat{a}_{0}^{\dagger}(\omega_{p1}+\omega_{p2}-\omega,L)$. We neglect the reverse process of converting the pairs back to pump photons, which is equivalent to neglecting double-pair emissions in SPDC or standard FWM. This approximation causes deviation less than $2\times10^{-6}$ for conversion efficiency $\eta=0.1$\%, and less than 0.02 for a single photon conversion efficiency up to $\eta=10$\%, as discussed in Appendix C. A treatment without the low gain approximation would allow simulation of Rabi oscillations between the single photon and photon pair, as required for the coherent photon conversion of Ref.~\cite{Langford2011Efficien}. 

\section{Single photon conversion efficiency}

The conversion efficiency of the single photon into a pair can now be derived from Eq.~(\ref{as_solved}). The spectral density of the photons created during the characteristic
time $T$ is given by 
\begin{equation}
n_\text{d}(\omega,L)=\left\langle \psi\left|\hat{a}_{0}^{\dagger}(\omega,L)\hat{a}_{0}(\omega,L)\right|\psi\right\rangle \label{eq:nd}.
\end{equation}
 The quantum state $\left|\psi\right\rangle $ is the input
state of the weak pump and generated photon pairs. 
For a single photon on pump 2, $\ket{\psi} =\ket{1} _{p2}\ket{0}_s\ket{0}_i$, where we label the lower frequency half of the output pair spectrum \emph{idler}, and the higher half \emph{signal.} The total number of photons generated during $T$ is then given by the integral of the spectral density over the output spectrum.

Putting Eq.~(\ref{as_solved}) into Eq.~(\ref{eq:nd}) gives the photon number spectral density per characteristic time 
\begin{eqnarray}
n_\text{d}(\omega,L) & = & \frac{T}{2\pi}\times4\gamma^{2}P_{1}P_{2}L^{2}\text{sinc}^{2}\left(\frac{KL}{2}\right)\label{eq:ns(omega)}\\
 & = & 4\gamma^{2}P_{1}\frac{\hbar\omega_{p2}}{2\pi}L^{2}\text{sinc}^{2}\left(\frac{KL}{2}\right)\nonumber,
\end{eqnarray}
where the generation of a photon at frequency $\omega$ implies the generation of its pair photon at frequency $\omega_{p1}+\omega_{p2}-\omega$. Let us now find the total number of photon pairs generated out of a single photon in cases of pulsed and continuous-wave pumping.

\subsection{Regime with both pumps pulsed}

If both pumps are pulsed simultaneously with a spectral width $\delta\omega_p$ and $T=\frac{2\pi}{\delta\omega_{p}}$, the total number of photon pairs generated per time $T$ (or per pulse for transform-limited pulses) is 

\begin{eqnarray}
N_{\text{pairs/pulse}} & = &\frac{1}{2} \int d\omega n_\text{d}(\omega,L)\label{eq:Pquasimc}\\
 & = & 4\gamma^{2}P_{1}P_{2}L^{2}\frac{\Delta\omega_{s}}{\delta\omega_{p}}\nonumber \\
 & = & 4\gamma^{2}P_{1}\frac{\hbar\omega_{p2}}{2\pi}L^{2}\Delta\omega_{s},\nonumber 
\end{eqnarray}
with 
\begin{equation}
\Delta\omega_{s}=\frac{1}{2}\int d\omega\text{sinc}^{2}\left(\frac{K(\omega_{p1},\omega_{p2},\omega)L}{2}\right) \label{eq:deltaomegas}
\end{equation} and $P_2=\frac{\hbar \omega_{p2}}{T}$,
where the factor 1/2 in the first line is due to the spectrum covering both signal and idler frequencies, leading to double-counting. The integral is over the whole spectral range except the two pump frequencies $\omega_{p1,p2}$.

The number of generated photons pairs per second is thus 
\begin{eqnarray}
N_{\text{pairs/sec}} & = & f_{\text{rep}}\times4\gamma^{2}P_{1}P_{2}L^{2}\frac{\Delta\omega_{s}}{\delta\omega_{p}}\nonumber \\
 & = & 4\gamma^{2}P_{\text{1avg}}P_2L^{2}\frac{\Delta\omega_{s}}{2\pi},\label{eq:Npulsedpulsed}
\end{eqnarray}
with $P_{1}=\frac{P_{\text{1avg}}}{f_{\text{rep}}}\times\frac{\delta \omega_p}{2\pi}$, where $f_{\text{rep}}$ is the repetition rate of the source.

\subsection{Regime with one pump pulsed and the other continuous-wave}

If one of the pumps is pulsed and the other is continuous-wave (CW), the output
photons will behave as if both pumps were pulsed at the repetition rate of
the pulsed one, which removes the necessity for time alignment. Taking the single photon pump as pulsed and the strong pump
as CW, we have $P_{1}=P_\text{{1avg}}$ and $P_{2}=\frac{\hbar\omega_{p2}}{T}$, which gives

\begin{eqnarray}
N_{\text{pairs/sec}}=f_{\text{rep}}\times4\gamma^{2}P_{\text{1avg}}P_{2}L^{2}\frac{\Delta\omega_{s}}{\delta\omega_{p}}. \label{eq:pulsed/cw}
\end{eqnarray}
The pair generation is independent of the single photon pulse
duration, depending only on its repetition rate and the strong laser's CW pump power. It is less efficient
by a factor $\frac{ f_{\text{rep}}}{\delta\omega_p}$ compared to when both pumps are pulsed.

If we want the weak pump to be CW, we can
argue an ``equivalent single photon'' pumping such that each pulse of the
strong pump sees on average one photon of the weak pump. Then we have to take $P_{2}=P_\text{{2avg}}=\frac{\hbar\omega_{p2}}{T}$, and

\begin{eqnarray}
N_{\text{pairs/sec}} & = & 4\gamma^{2}P_{\text{1avg}}P_{\text{2avg}}L^{2}\frac{\Delta\omega_{s}}{2\pi}.\label{eq:Pquasimc-2-1}
\end{eqnarray}

This generation is equivalent to the pulsed/pulsed pumping, which is
not surprising since the ``equivalent single photon'' pumping
is CW pumping with the same peak power as the pulsed pumping. This means many more photons of pump p2 enter the fiber, but only the ones which overlap a strong pump pulse can convert into pairs.

\subsection{Regime with two continuous-wave pumps}

With an input made out of a weak continuous pump 2 and a strong continuous pump 1, then the number
of photons generated per second is straightforward, 

\begin{eqnarray}
N_{\text{pairs/sec}} & = & 4\gamma^{2}P_\text{{1avg}}P_\text{{2avg}}L^{2}\frac{\Delta\omega_{s}}{2\pi},\label{eq:Pquasimc-1-1}
\end{eqnarray}
however it is not obvious to define what qualifies as a single photon for a CW pump.
This regime can reach the same efficiency as the pulsed/pulsed case if either of the pumps' CW average power is raised to the peak power of the pulsed/pulsed case. This would be difficult in practice, as peak powers can be four orders of magnitude larger than average powers for the example of modelocked picosecond lasers.

\section{Candidate fibers for maximizing conversion efficiency}\label{sec:pm}

In this section we compare three fiber types with unique methods of phasematching to find the best for single photon conversion. The single photon conversion efficiency can be defined, for a weak pumping field composed of $N_{\text{p2}}$ photons per time unit, as $\eta =  \frac{N_{\text{pairs}}}{N_{\text{p2}}}$, {$N_{\text{pairs}}$ being the number of photon pairs generated during the same time unit. 
If the weak pumping field is a single photon, the pulsed/pulsed configuration gives the highest conversion efficiency for a given input average power of pump 1 (compare Eq.~(\ref{eq:Npulsedpulsed})  with Eqs.~(\ref{eq:pulsed/cw}) and (\ref{eq:Pquasimc-1-1})). 
Let us consider strategies for maximizing the single photon conversion in this regime. When $N_{\text{p2}}=1$, the conversion efficiency as given by Eq.~(\ref{eq:Pquasimc}) is

\begin{eqnarray}
\eta & = & \frac{N_{\text{pairs}}}{N_{\text{p2}}}=4\gamma^{2}P_{1}\frac{\hbar\omega_{p2}}{2\pi}L^{2}\Delta\omega_{s}.\label{eq:efficiency} 
\end{eqnarray}
The parameters that can be tuned to maximize conversion efficiency are nonlinearity $\chi^{(3)}$
and mode area $A_{\text{eff}}$ through $\gamma$, length $L$, phasematching bandwidth $\Delta\omega_{s}$, and peak power $P_{1}$ of the strong pump. Since $\gamma$ is squared, decreasing $A_{\text{eff}}$ and increasing $\chi^{(3)}$
will have the greatest effect. By contrast, conversion efficiency appears quadratic in the length of fiber, $L$, but the signal and idler bandwidths given by Eq.~~(\ref{eq:deltaomegas}) will vary approximately with $1/L$, giving an overall linear dependence on fiber length.
 The spectral width $\Delta\omega_{s}$ can also vary independent of $L$ from tiny ($\delta\omega$) to hundreds of nano-meters, depending
on the pump configuration, and most importantly on the type of phasematching chosen. 

We examine three candidates for maximizing conversion efficiency, corresponding to the three main methods of phasematching in optical fibers: birefringence, operation near a
zero dispersion wavelength (ZDW), and nonlinear phasematching using self-phase modulation. The phasematched
frequencies generated by the use of birefringence are spectrally narrow and highly tunable. The frequencies
phasematched around the ZDW or due to nonlinear phase modulation
can have a broader spectrum, and are centered around or near the ZDW. We compare the potential for single photon conversion in three different fiber types corresponding to those three types of phasematching, and find the optimal parameters to maximize pair generation.

The phase mismatch can be expressed as a Taylor expansion around the central frequency $\omega_0$ as
\begin{equation}
K(\Omega)=\beta_{2}(\omega_{0})\left(\Omega^{2}-\Delta\omega^{2}\right)+\frac{\beta_{4}(\omega_{0})}{12}\left(\Omega^{4}-\Delta\omega^{4}\right)+\gamma P_{1}\label{eq:K_non_degen},
\end{equation}
with the central frequency $\omega_0 = \frac{\omega_{p2}+\omega_{p1}}{2}$, the offset frequency $\Omega=\omega-\omega_{0}$, the pump offset $\Delta\omega=\frac{\omega_{p2}-\omega_{p1}}{2}$, and dispersion coefficients given by
\begin{equation}
\beta_{i}(\omega_{0})=\left(\frac{\partial^{i}\beta(\omega)}{\partial\omega^{i}}\right)_{\omega_{0}}.
\end{equation}

 Two schemes can be considered when the pump wavelengths are nondegenerate as required for single-photon FWM: external pumping,
with generation of new wavelengths in between the pump wavelengths,
or internal pumping, with generation of new wavelengths to the exterior.
We focus on external pumping as illustrated in Fig.~\ref{fig:General-wavelength-configuration} because, assuming the strong pump has the highest wavelength, it allows filtering the main Raman noise from the strong pump as this will be at higher wavelengths still. However, the large separation in pump wavelengths can lead to temporal walk-off between the pump pulses in the fiber, reducing efficiency. This effect is mitigated by situating the pumps symmetrically about the ZDW.

\begin{figure}
\includegraphics[width=\columnwidth]{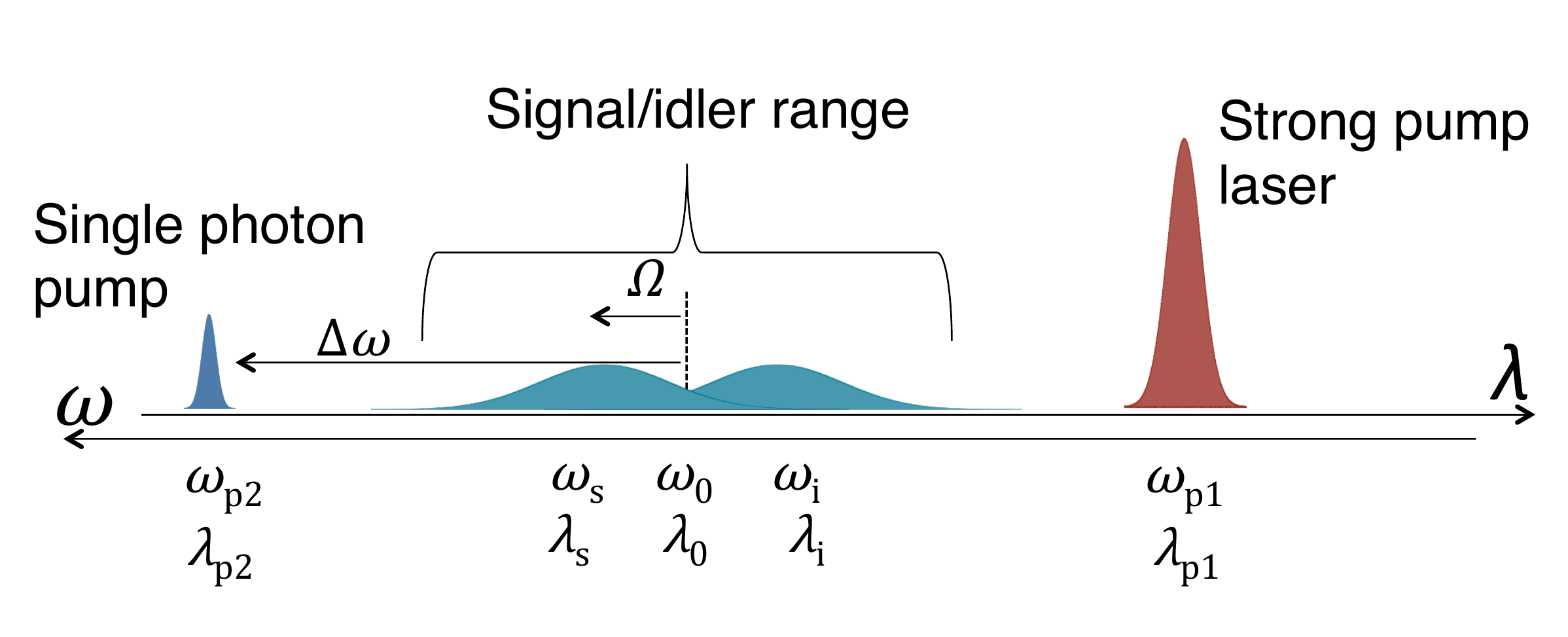}
\caption{[Color online] Arrangement of pump and signal/idler wavelengths (amplitudes and widths not to scale). The main source of noise, spontaneous Raman scattering from the strong pump, will occur to the far right of this figure, allowing its removal by spectral filtering. \label{fig:General-wavelength-configuration}}
\end{figure}

\subsection{Polarization maintaining fiber: birefringent phasematching}

Standard polarization maintaining (PM) fiber exhibits a birefringence
large enough to achieve phasematching some
dozens to hundreds of tera-hertz from the pumps ($\sim 100$~nm)~\cite{Smith:09}. These fibers are commercially available, with lengths up to kilometers, and spatially uniform. The phasematching is easy to obtain and widely tunable
by tuning the pump wavelengths. Further, the birefringent phasematching means that the photon pairs can come out with opposite polarizations from the pumps, enabling polarization filtering of the pumps and associated Raman noise. However, the relatively large core size leads to a modest waveguide nonlinear parameter of $\gamma =4.6 \times10^{-3} $~W$^{-1}\cdot$m$^{-1}$ in our example below.

We consider the two pumps co-polarized along
the fast axis and the generated signal and idler polarized along the
slow axis, which gives total phase mismatch 

\begin{eqnarray}
K(\Omega)&=&\beta_{2}(\omega_{0})\left(\Omega^{2}-\Delta\omega^{2}\right)+\frac{\beta_{4}(\omega_{0})}{12}\left(\Omega^{4}-\Delta\omega^{4}\right)\nonumber\\
& &+\gamma P_{1}+2\omega_{0}\frac{\delta n}{c},\label{eq:pmBiref}
\end{eqnarray}
where the birefringence $\delta n=n^{slow}- n^{fast}$ is written separately from the dispersion coefficients.
Far from the ZDW ($\beta_2\gg\beta_4,\gamma P_1$), the
phase matched frequencies are
\begin{equation}\label{eq:Omega0_BR}
\Omega^{2}=-\frac{2\omega_{0}}{\beta_{2}(\omega_{0})}\frac{\delta n}{c}+\Delta\omega^{2}.
\end{equation}

We consider a silica PM fiber with birefringence $\delta n = 3\times10^{-4}$ (e.g. Panda PM630), and take both pumps pulsed with 80~MHz repetition rate and 5~ps pulses. We take a 5~W average power for the strong pump, and a single photon for the weak pump. The walk-off length between the two pump pulses in this configuration being 18 cm, we consider an 11 cm fiber which gives an effective interaction length of $L=$ 10 cm.

With the strong pump at 890~nm and the weak pump at 660~nm, we obtain a signal and idler phase
matched at 728~nm and 790~nm with spectral width $\Delta\omega_{s}=7$~
rad THz (2 nm), as shown in Fig.~\ref{fig:PM signal plot}. The conversion efficiency given by Eq.~(\ref{eq:efficiency}) is 
 $\eta=2\times10^{-8}$, well below that achievable in $\chi^{(2)}$ media.  We plot the signal and idler spectral density (photons per (rad Hz) per pulse) in Fig.~\ref{fig:PM signal plot}, accurate to the precision of our frequency space mapping, $\delta\omega_p = 1.3 $~rad~THz (0.5~nm, given by the width of the grey lines on the graph). 
\begin{figure}[htp] 
\includegraphics[width=\columnwidth]{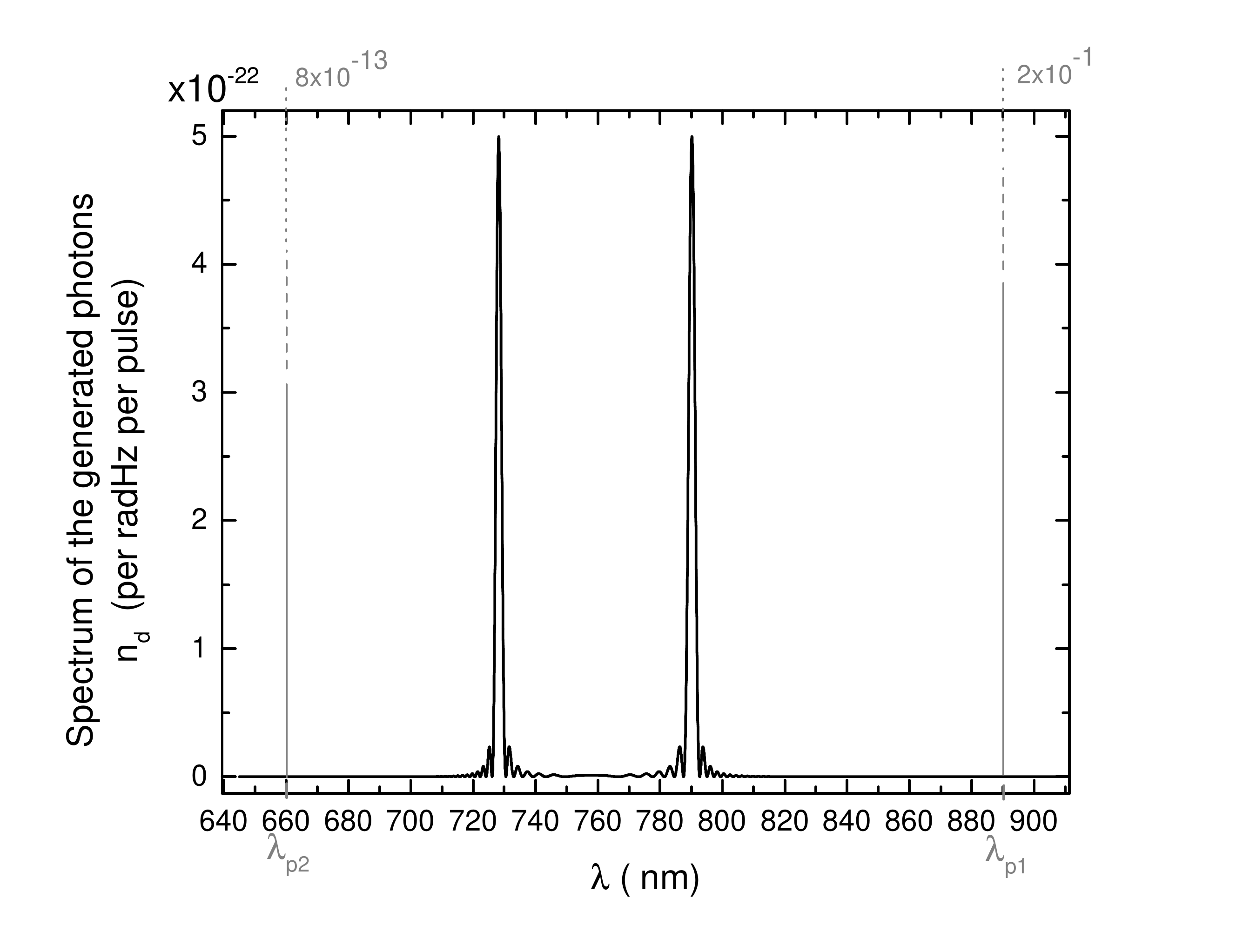}\caption{Converting a single photon to a pair via FWM is possible in principle using a PM fiber, but the very narrow phasematching limits the efficiency to $\eta=2\times10^{-8}$ in this example. The quasi-monochromatic pump wavelengths are represented by grey lines of width 1.3 rad~THz with values labeled above the graph (well above the y-axis shown), while the generated signal and idler spectra are the black lines in the centre.\label{fig:PM signal plot}}
\end{figure}

\subsection{Microstructured fiber: phasematching near the zero-dispersion wavelength}

Phasematching occurs in a fiber near the ZDW when the material and waveguide
contributions to dispersion cancel. We will take the example of silica microstructured fibers, which are commercially available
and can be fabricated to exhibit a ZDW
in the visible and telecom ranges. The interest in such a fiber is that the core
can be much smaller than regular single mode fibers, thus increasing the waveguide nonlinear parameter, e.g. up to $\gamma=2.7\times10^{-2}$ W$^{-1}\cdot$m$^{-1}$ in our example, with lengths up to a few meters~\cite{Rarity:05}. The spectral broadness of the phasematching
depends on the length considered, and is only tuneable in a small range once the ZDW is chosen.

We model a microstructured fiber with core diameter of 1.8~$\mu$m and air fraction 0.72 in the cladding, which give the ZDW at 716~nm. The wavelengths and the pump powers are altered slightly from the previous example to achieve phasematching. We take a 1~W average power for the strong pump
in a 2~m long fiber, with 2~ps long pump pulses and 2~ps long single photon
pulses at 80~MHz repetition rate. As a consequence of working near the ZDW, the walkoff length is now over 100~m, since the pumps have approximately the same propagation constant $\beta$ on either side of the ZDW. The single photon frequency
is at wavelength 676.75~nm and the strong pump is now at 760~nm.
Simulations give a much broader spectrum for the signal and idler (around 160 rad THz, Fig.~\ref{fig:PM signal plot-1}), and consequently the
efficiency, still given by Eq.~(\ref{eq:efficiency}), is now up to $\eta=4\times10^{-4}$. Even including filtration of the generated photons between 686 nm and 750 nm to allow pump removal, this efficiency four orders of magnitude better than the PM fiber. 

\begin{figure}
\includegraphics[width=\columnwidth]{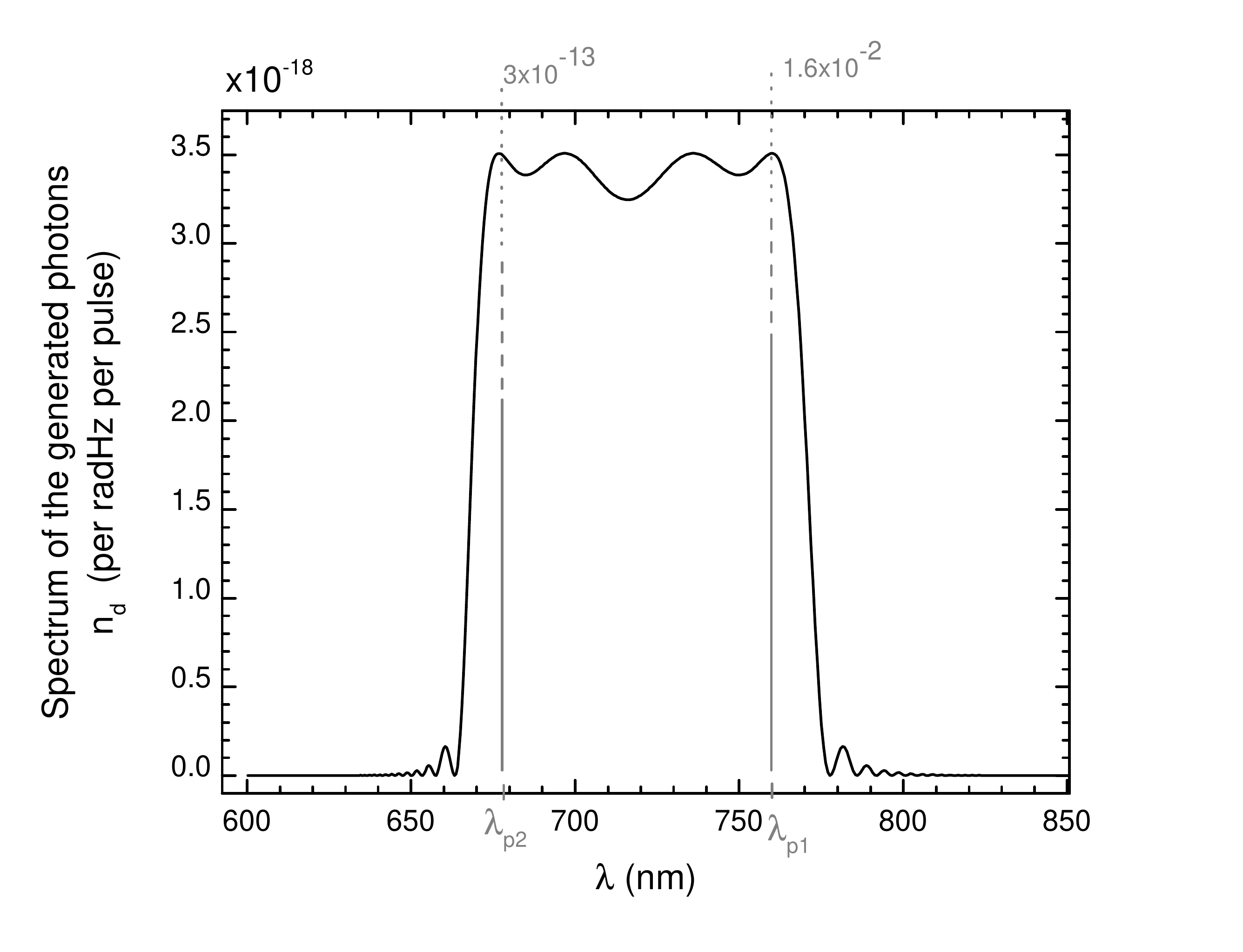}\caption{ A much broader signal and idler spectrum is obtained near the ZDW using a microstructured fiber. Even after filtering between 675~nm and 760~nm to remove noise photons near the pumps, the efficiency $\eta=4\times10^{-4}$ over the remaining signal/idler range is four orders of magnitude larger than for the PM fiber. The grey lines represent the pumps' wavelengths, widths (now 3.1 rad THz) and intensity as in Fig. \ref{fig:PM signal plot}. \label{fig:PM signal plot-1}}
\end{figure}

However, this method of phasematching is very sensitive to core diameter and pump wavelengths.   For example, a 0.5~nm deviation of pump wavelength changes the output spectral shape completely and, even if still phasematched, may give no pairs at the center of the spectrum. Additionally, obtaining a 2~m microstructured fiber with good uniformity for the whole length is not straightforward, as some variations in the core diameter will occur that deteriorate the perfect phasematching.

\subsection{Chalcogenide microwire fiber: phasematching due to self-phase modulation}

Achieving the best conversion efficiency requires ultrahigh nonlinearity and small cross-sectional area. These can be achieved by tapering fibers made of chalcogenide glass as in Refs. \cite{Rochette-fabrication,Rochette-FWM}. The chalcogenide As$_2$Se$_3$ has $\chi^{(3)}$ three orders of magnitude larger than that of silica glass and core diameters in the tapered microwire region can be as small as 500~nm thanks to its large refractive index, while still maintaining good coupling to standard single mode fiber and lengths beyond 10~cm. These microwires exhibit ultrahigh waveguide nonlinear parameters up to $\gamma=180$~W$^{-1}\cdot $m$^{-1}$.

This large $\gamma$ directly leads to high conversion efficiency, but also allows nonlinear phasematching. As seen in Eq.~(\ref{eq:K_non_degen}) it is possible to compensate
for positive or negative linear phase mismatch by the nonlinear contribution $\gamma P_1$ 
due to the strong pump self-phase modulation. The higher the dispersive mismatch, the higher the pump powers must be to compensate, so moderate pump powers still require working near the ZDW.
In the external pumping configuration of Fig.~\ref{fig:General-wavelength-configuration}, the dispersion $\beta_2$ or $\beta_4$ has to
be positive to compensate for self-phase modulation because the pump offset is greater than the frequency offset, i.e. $\Delta\omega^2>\Omega^2$. 
The pump power necessary
to reach perfect phasematching is 
\begin{equation}
P_{1}=\frac{1}{\gamma}\left(\beta_{2}(\omega_{0})\Delta\omega^{2}+\frac{\beta_{4}(\omega_{0})}{12}\Delta\omega^{4}\right)\label{eq:Pfonctionbeta}.
\end{equation}
 
For convenient all-telecom operation and to avoid the two-photon absorption at short wavelengths in chalcogenide glass As$_2$Se$_3$~\cite{Lenz2000-TPA}, we take the example of the FWM scheme pumped at 1480~nm by the single photon and at 1620~nm by a strong pump. The two fields are pulsed at 80~MHz with 2~ps long pulses. The fiber considered
is similar to the samples described in Ref.~\cite{Rochette-FWM}.
A fiber diameter of $0.555\,\mu$m for the microwire gives a dispersion
coefficient at 1550~nm of $\beta_{2}(\omega_{0})=0.05$~ps$^2$/m, and $\beta_4$ is negligible. Phasematching is achieved for a 0.8~W peak power, which corresponds
to an average power of only 0.13 mW. Simulation of the spectral
density is given in Fig.~\ref{fig:Chalco signal} in a 10~cm long microwire section, where the walk-off length between the two pump pulses is now so large as to be effectively infinite. Both the high intrinsic
$\chi^{(3)}$ of the chalcogenide and the strong confinement allows
to reach a conversion efficiency of $\eta=1.1\times10^{-3}$. However, as in the silica microstructured fibers, caution must be taken in filtering the desired photons, since high nonlinearity means high Raman noise, large phase modulation broadening and other undesired interactions such as degenerate FWM from the strong pump.

\begin{figure}
\includegraphics[width=\columnwidth]{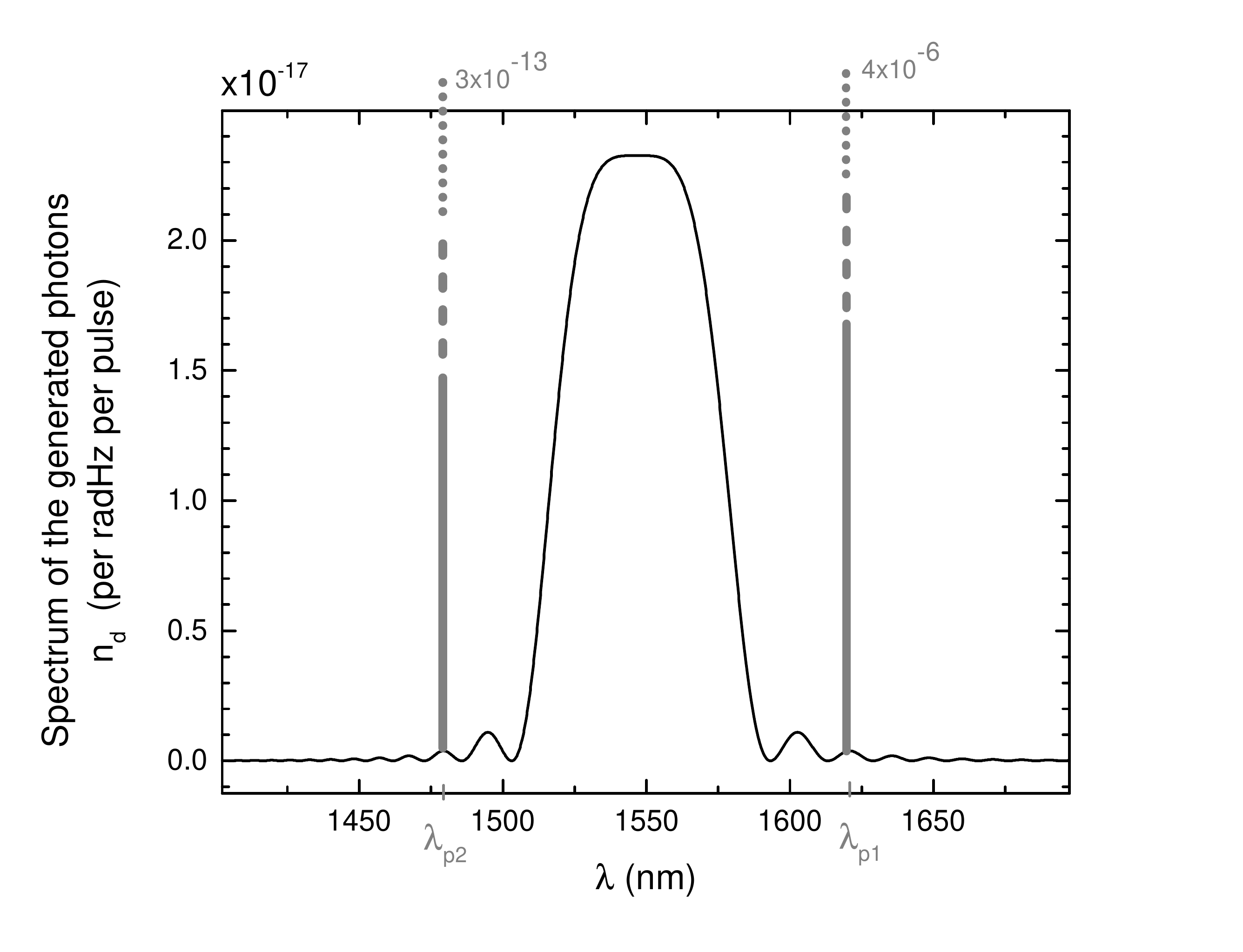}\caption{Due to high intrinsic  $\chi^{(3)}$ and strong confinement, the chalcogenide microwire gives the highest conversion rate, with an efficiency of $\eta=1.1\times10^{-3}$. The grey lines represent the pumps' wavelengths, widths (3.1~rad~THz) and intensity as in Fig. \ref{fig:PM signal plot}. \label{fig:Chalco signal}}
\end{figure}

\subsection{Experiments}

Tests in our lab have shown that the nonlinear interaction in standard birefringent fibers is indeed very weak, making them unsuitable for single photon conversion. The implementation in Ref.~\cite{Langford2011Efficien} in birefringent microstructured fibers allowed an inferred conversion efficiency of $3\times10^{-9}$, with the weak pump kept at 4.8~$\mu$W effective average power, well above the single-photon level. This was achieved for strong pump average powers under 100~mW and employed narrowband birefringent phasematching, limiting conversion efficiency, but demonstrating the principle of single photon conversion. Finally, we have performed preliminary experiments on the chalcogenide microwires, verifying nondegenerate, external pumping phasematching conditions very similar to those shown above.

\section{Conclusion}

\begin{table}
\begin{ruledtabular}
\begin{tabular}{l c c c}
Fiber  & Average pump 		&         Conversion   		& Photon pairs     \\ 
   type                    & power (mW)   		& efficiency $\eta$ 		& per second \\ \hline
Biref. (silica)    			& 5000 		& $2\times10^{-8}$   	& 1.6  \\
Microstr. (silica)		& 1000	& $4\times10^{-4}$   	& 32,000   \\
Microwire (As$_2$Se$_3$)    		&0.13 		& $1\times10^{-3}$	 	& 80,000 \\ 
\end{tabular}
\end{ruledtabular}
\caption{Summary of expected single photon to pair conversion efficiency and strong pump power required for the three fiber types considered: birefringent silica fibers, microstructured silica fibers, and chalcogenide As$_2$Se$_3$  microwired fibers. Considering the 80~MHz repetition rate and inputting one photon per pulse, we also calculate the number of pairs produced per second.}
\label{tab:summary}
\end{table}

We have predicted a promising result in the conversion of single photons into pairs via four-wave mixing. As shown from our simulations based on evolution of the quantum field operators, conversion efficiencies up to 0.1\% should be achievable in chalcogenide microwires. The results in the three types of fibers we modeled are summarized in Table~\ref{tab:summary}. 

In this work, the low gain approximation is sufficient for applications in generating large entangled states and photon heralding, though a non-perturbative approach keeping all orders of gain would make for an interesting study and allow exploring the deterministic pair generation $\ket{1}_{p2}\rightarrow\ket{11}_{si}$ and oscillatory $\ket{1}_{p2}\rightarrow\ket{11}_{si}\rightarrow\ket{1}_{p2}\rightarrow\ldots$ regimes. However, finding a material enabling photon conversion with an efficiency high enough to justify this non-perturbative approach remains a challenge.

Interestingly, neither the derivation nor value of the conversion efficiency we found depend on the single photon being quantized: the result can equally be obtained by assuming a classical pulse with the same input peak power as the single photon. This implies there is no new quantumness in this process, beyond the well-established spontaneous generation of pairs in standard spontaneous FWM or SPDC.

 In implementations, caution must be paid to the various sources of noise possible: degenerate FWM and spontaneous Raman scattering~\cite{Clark:12} from the strong pump, and even second orders or combinations of these effects. 

\begin{acknowledgments}
We thank Piotr Kolenderski for helpful comments and acknowledge support from NSERC (CGS, Discovery, CREATE, RTI), Ontario Ministry of Research and Innovation, CIFAR, FedDev Ontario, Industry Canada, and CFI. The authors are thankful to Coractive High-Tech for providing the chalcogenide glass used in the experiments.
\end{acknowledgments}

\appendix
\section{Derivation of momentum generators}\label{app:momentum}

The two pumps are considered monochromatic or quasi-monochromatic,
perfectly overlapping, of identical pulse duration and spectral width $\delta\omega$. The quantization time is chosen as the transform-limited pulse duration, $T=2\pi/\delta\omega$. 

The quantum field operator can be written in the continuous limit as 

\begin{eqnarray}
\hat{E}(z,t)  = \int d\omega\sqrt{\Omega(\omega)}\hat{a}(\omega,z)e^{-i\ensuremath{\omega}t}+h.c.
\end{eqnarray}
where the integral runs from zero to infinity, with the notation simplified by the introduction of the variable $\Omega(\omega)=\frac{\mbox{\ensuremath{\hbar}}\omega}{4\pi\varepsilon_{0}cA_{eff}n(\omega)}$.

The momentum operator is given by Eq.~(\ref{eq:g}), from which we can derive its linear part $\hat{G}_l(z)$ and nonlinear part $\hat{G}_{nl}(z)$.

\begin{eqnarray*}
\hat{G}_{l}(z) & = & \int_{A_{eff}}dS\int_{0}^{+T}dt\iint d\omega d\omega'\varepsilon_{0}\left(\chi^{(1)}(\omega)+1\right)\nonumber\\
& & \times\sqrt{\Omega(\omega)}\hat{a^\dagger}(\omega,z)e^{i\ensuremath{\omega}t}\sqrt{\Omega(\omega')}\hat{a}(\omega',z)e^{-i\ensuremath{\omega'}t}+h.c.\\
 & = & 2A_{eff}\iint d\omega d\omega'\varepsilon_{0}n(\omega)^{2}\Omega(\omega)\hat{a}^\dagger(\omega,z)\hat{a}(\omega,z)\nonumber\\
 & &\times2\pi\delta(\omega-\omega')\\
 & = & 4\pi A_{eff}\varepsilon_{0}\int d\omega n(\omega)^{2}\Omega(\omega)\hat{a}^\dagger(\omega,z)\hat{a}(\omega,z)\\
 & = & \int d\omega\frac{\mbox{\ensuremath{\hbar}\ensuremath{\omega}}}{c}n(\omega)\hat{a}^\dagger(\omega,z)\hat{a}(\omega,z)
\end{eqnarray*}
where we used $\intop_{0}^{T}dte^{i(\omega-\omega')t}=2\pi\delta(\omega-\omega')$ (since the integration time matches the quantization time). The linear evolution of any annihilation operator can thus be deduced as in Eq.~(\ref{eq:a_lin}).
The nonlinear momentum operator evolves according to 

\begin{equation}
\hat{G_{nl}}(z)  =  \int_{A_{eff}}dS\int_{0}^{+T}dt\hat{P_{nl}}^{(-)}(z,t)\hat{E}^{(+)}(z,t)+h.c.\label{eq:g-1}.
\end{equation}
If we only consider FWM as generating propagating modes, the relevant nonlinear polarization is $\hat{P_{nl}}(z,t)=\varepsilon_{0}\chi^{(3)}\vdots\hat{E}^{3}(z,t)$. We can decompose
$\hat{G_{nl}}(z)$ into two parts, one for FWM and the other for phase modulation, as
$\hat{G}_{nl}(z)=\hat{G}_{nl}^{FWM}(z)+\hat{G}_{nl}^{PhMod}(z)$. Then
\begin{eqnarray}
 &  & \hat{G}_{nl}(z)=\nonumber \\
 &  & \varepsilon_{0}\chi^{(3)}\int_{A_{eff}}dS\int_{0}^{+T}dt\,\left[\int d\omega\sqrt{\Omega(\omega)}\hat{a}^\dagger(\omega,z)e^{i\ensuremath{\omega}t}+h.c.\right]\nonumber\\
 & &\times\left[\int d\omega'\sqrt{\Omega(\omega')}\hat{a}^\dagger(\omega',z)e^{i\ensuremath{\omega'}t}+h.c.\right]\nonumber \\
 &  & \times\left[\int d\omega''\sqrt{\Omega(\omega'')}\hat{a}^\dagger(\omega'',z)e^{i\ensuremath{\omega''}t}+h.c.\right]\nonumber\\
 & &\times\left[\int d\omega'''\sqrt{\Omega(\omega''')}\hat{a^\dagger}(\omega''',z)e^{i\ensuremath{\omega'''}t}+h.c.\right].\label{eq:Gnl_general}
\end{eqnarray}
We keep only the frequencies that will propagate in the fiber: the frequencies around the two pumps, and frequencies
generated by FWM with these 2 pumps. We can separate the operators into four frequency parts

\begin{eqnarray}
 &  & \int d\omega\sqrt{\Omega(\omega)}\hat{a}(\omega,z)e^{-i\ensuremath{\omega}t}=\label{eq:E expand}\\
 &  & \int_{\Delta\omega_{p1}}d\omega_{1}\sqrt{\Omega(\omega_{1})}\hat{a}(\omega_{1},z)e^{-i\ensuremath{\omega_{1}}t}\nonumber\\
 & &+\int_{\Delta\omega_{p2}}d\omega_{2}\sqrt{\Omega(\omega_{2})}\hat{a}(\omega_{2},z)e^{-i\omega_{2}t}\nonumber \\
 &  & +\int_{\Delta\omega_{s}}d\omega\sqrt{\Omega(\omega)}\hat{a}(\omega,z)e^{-i\ensuremath{\omega}t}\nonumber \\
 &  &+\int_{\Delta\omega_{i}}d\omega'\sqrt{\Omega(\omega')}\hat{a}(\omega',z)e^{-i\ensuremath{\omega'}t},\nonumber 
\end{eqnarray}
where  $\Delta\omega_{s}$ and $\Delta\omega_{i}$
are a wide frequency range around the central frequencies of the photon pairs generated by FWM and $\Delta\omega_{p1,p2}$ are a wide range frequency around the two pump wavelengths. Since we assumed our two pumps are monochromatic or quasi-monochromatic and took their bandwidth as
the frequency step $\Delta\omega_{p1}=\Delta\omega_{p2}=\delta\omega=2\pi/T$, we can write
$\int_{\Delta\omega_{p1}} d\omega_{1}\sqrt{\Omega(\omega_{1})}\hat{a}(\omega_{1},z)=\delta\omega\sqrt{\Omega(\omega_{p1})}\hat{a}(\omega_{p1},z)=\frac{2\pi}{T}\sqrt{\Omega(\omega_{p1})}\hat{a}(\omega_{p1},z)$
with $\omega_{p1}$ the central frequency of pump 1, and the same
for pump 2. For more clarity in the expressions with respect to the other operators, we keep
the pump mode operators dimensioned as $\left[\hat{a}(\omega_{p1,p2},z)\right]=\sqrt{\frac{1}{\delta\omega}}$,
so the number of pump photons travelling through a plane
of position $z$ during the time interval $T$ is $\frac{2\pi}{T}\left\langle \hat{a}^\dagger(\omega_{p1,2},z)\hat{a}(\omega_{p1,2},z)\right\rangle $. 
Then we can write
 
\begin{eqnarray}
 &  & \int d\omega\sqrt{\Omega(\omega)}\hat{a}(\omega,z)e^{-i\ensuremath{\omega}t}=\label{eq:E expand2}\\
 &  & \frac{2\pi}{T}\left[ \sqrt{\Omega(\omega_{p1})}\hat{a}(\omega_{p1},z)e^{-i\ensuremath{\omega_{p1}}t}+\sqrt{\Omega(\omega_{p2})}\hat{a}(\omega_{p2},z)e^{-i\omega_{p2}t}\right] \nonumber \\
 &  & +\int_{\Delta\omega_{s}}d\omega\sqrt{\Omega(\omega)}\hat{a}(\omega,z)e^{-i\ensuremath{\omega}t}\nonumber \\
 & & +\int_{\Delta\omega_{i}}d\omega'\sqrt{\Omega(\omega')}\hat{a}(\omega',z)e^{-i\ensuremath{\omega'}t}\nonumber.
\end{eqnarray}

The FWM part of the nonlinear momentum is then 
\begin{widetext}
\begin{eqnarray}
 &  & \hat{G}_{nl}^{FWM}(z)= 24\times\left(\frac{2\pi}{T}\right)^{2}\varepsilon_{0}\chi^{(3)}\int_{A_{eff}}dS\int_{0}^{+T}dt\,\int_{\Delta\omega_{s}}d\omega\sqrt{\Omega(\omega)}\hat{a}^\dagger(\omega,z)\nonumber\\
&&\times\int_{\Delta\omega_{i}}d\omega'\sqrt{\Omega(\omega')}\hat{a}^\dagger(\omega',z)\sqrt{\Omega(\omega_{p1})}\hat{a}(\omega_{p1},z)\sqrt{\Omega(\omega_{p2})}\hat{a}(\omega_{p2},z)e^{i\ensuremath{\Delta\omega}t}+h.c.\nonumber 
\end{eqnarray}
or, if we write the operators as product of their linear and nonlinear parts,

\begin{eqnarray}
 &  & \hat{G}_{nl}^{FWM}(z)=24\times\left(\frac{2\pi}{T}\right)^{2}\varepsilon_{0}\chi^{(3)}\int_{A_{eff}}dS\int_{0}^{+T}dt\,\int_{\Delta\omega_{s}}d\omega\sqrt{\Omega(\omega)}\hat{a}_{0}^\dagger(\omega,z)\nonumber \\
 &  & \times\int_{\Delta\omega_{i}}d\omega'\sqrt{\Omega(\omega')}\hat{a}_{0}^\dagger(\omega',z)\sqrt{\Omega(\omega_{p1})}\hat{a}_{0}(\omega_{p1},z)\sqrt{\Omega(\omega_{p2})}\hat{a}_{0}(\omega_{p2},z)e^{i\ensuremath{\Delta\omega}t}e^{-i\Delta kz}+h.c..\nonumber 
\end{eqnarray}

The factor 24 comes from all the possible combinations of the mode
operators. Here $\Delta\omega=\omega+\omega'-\omega_{p1}-\omega_{p2}$
and $\Delta k=\beta(\omega)+\beta(\omega')-\beta(\omega_{p1})-\beta(\omega_{p2})$.
Using again $\intop_{0}^{T}dte^{i\Delta\omega t}=2\pi\delta(\Delta\omega)$ and $\int d\omega'\delta(\Delta\omega)=1$, and evaluating the cross-sectional area integral, we
have 

\begin{eqnarray}
\hat{G}_{nl}^{FWM}(z) & = & 24\times2\pi\times\left(\frac{2\pi}{T}\right)^{2}\varepsilon_{0}\chi^{(3)}A_{eff}\sqrt{\Omega(\omega_{p1})}\sqrt{\Omega(\omega_{p2})}\int_{\Delta\omega_{s}}d\omega\sqrt{\Omega(\omega)}\sqrt{\Omega(\omega_{p1}+\omega_{p2}-\omega)}\times\nonumber \\
 &  & \hat{a}_{0}^\dagger(\omega,z)\hat{a}_{0}^\dagger(\omega_{p1}+\omega_{p2}-\omega,z)\times\hat{a}_{0}(\omega_{p1},z)\hat{a}_{0}(\omega_{p2},z)e^{-i\Delta kz}+h.c..
\end{eqnarray}
\end{widetext}

We can extend the integral over the signal over the whole spectrum except
the two pumps' frequencies and add a factor
1/2 for double-counting signal and idler frequencies. Then 

\begin{eqnarray}
& &\hat{G}_{nl}^{FWM}(z)  =3\times\frac{2\pi}{T}  \chi^{(3)}\frac{\mbox{\ensuremath{\hbar}}^{2}}{\varepsilon_{0}c^{2}A_{eff}T}\sqrt{\frac{\omega_{p1}}{n(\omega_{p1})}\frac{\omega_{p2}}{n(\omega_{p2})}}\nonumber \\
 &  & \times\int d\omega\sqrt{\frac{\omega}{n(\omega)}}\sqrt{\frac{\omega_{p1}+\omega_{p2}-\omega}{n(\omega_{p1}+\omega_{p2}-\omega)}}\hat{a}_{0}^\dagger(\omega,z) \\
 &  & \times\hat{a}_{0}^\dagger(\omega_{p1}+\omega_{p2}-\omega,z)\hat{a}_{0}(\omega_{p1},z)\hat{a}_{0}(\omega_{p2},z)e^{-i\Delta kz}+h.c..\label{eq:Gnl_FWM_Co2}\nonumber
\end{eqnarray}

Now let's look for the phase modulation term. By definition of
phase modulation, we keep only the terms with no phase that arise
from the expansion (\ref{eq:Gnl_general}) of the nonlinear momentum.
We then obtain :

\begin{eqnarray}
& &\hat{G}_{nl}^{ph\, mod}(z)  =  \int_{A_{eff}}dS\,\varepsilon_{0}\chi^{(3)}\times\\
 &  & \left[6\times\left(\frac{2\pi}{T}\right)\sum_{k=s,i,p_{1},p_{2}}\left(\int_{\Delta\omega_{k}}d\omega\Omega(\omega)\hat{a}_{0}^\dagger(\omega,z)\hat{a}_{0}(\omega,z)\right)^{2}+\nonumber \right.\\
 &  & 12\times\left(\frac{2\pi}{T}\right)\sum_{k=s,i,p_{1},p_{2}}\sum_{j\neq k}\left(\frac{2\pi}{T}\right)\int_{\Delta\omega_{k}}d\omega\int_{\Delta\omega_{j}}d\omega'\nonumber\\
 & &\left. \Omega(\omega)\Omega(\omega')\hat{a}_{0}^\dagger(\omega,z)\hat{a}_{0}(\omega',z)\hat{a}_{0}^\dagger(\omega',z)\hat{a}_{0}(\omega,z)\right]\nonumber ,
\end{eqnarray}
and if we sum over all the frequencies in the integrals, the momentum generator
collapses into 
\begin{eqnarray}
& &\hat{G}_{nl}^{ph\, mod}(z)  =  \left(\frac{2\pi}{T}\right)\times2\pi\times A_{eff}\varepsilon_{0}\chi^{(3)}\times\\
 &  & \left[12\iint d\omega d\omega'\Omega(\omega)\Omega(\omega')\hat{a}_{0}^\dagger(\omega,z)\hat{a}_{0}(\omega',z)\hat{a}_{0}^\dagger(\omega',z)\hat{a}_{0}(\omega,z)\right.\nonumber \\
 &  & \left.-6\int d\omega \frac{2\pi}{T}\left(\Omega(\omega)\hat{a}_{0}^\dagger(\omega,z)\hat{a}_{0}(\omega,z)\right)^{2}\right]\nonumber
\end{eqnarray}
or
\begin{eqnarray}
& &\hat{G}_{nl}^{ph\, mod}(z)  =  3\chi^{(3)}\frac{\mbox{\ensuremath{\hbar}}^{2}}{\varepsilon_{0}c^{2}A_{eff}T}\\
 &  & \left[\iint d\omega d\omega'\frac{\omega}{n(\omega)}\frac{\omega'}{n(\omega')}\hat{a}_{0}^\dagger(\omega,z)\hat{a}_{0}(\omega',z)\hat{a}_{0}^\dagger(\omega',z)\hat{a}_{0}(\omega,z)\right.\nonumber \\
 &  & \left.-\frac{1}{2}\int d\omega \frac{2\pi}{T}\left(\frac{\omega}{n(\omega)}\hat{a}_{0}^\dagger(\omega,z)\hat{a}_{0}(\omega,z)\right)^{2}\right]\nonumber.
\end{eqnarray}

\section{Low gain approximation and derivation of mode operators}\label{app:BCH}
We can now derive the mode operators from Eq.~(\ref{eq:da/dz}),
for any frequency generated in the fiber, which gives Eqs.~(\ref{eq:da/dz_continue}), (\ref{eq:das/dz}), and (\ref{dap/dz}).

We proceed by first factoring out the phase modulation with the variable
change $\hat{a}_{0}(\omega,z)=\hat{a}_{0}'(\omega,z)e^{i2\gamma P_{1}z}$, which gives for the generated modes' FWM evolution
\begin{widetext}
\begin{eqnarray}
 &  & \frac{\partial \hat{a}_{0}'(\omega,z)}{\partial z}e^{i2\gamma P_{1}z}+i2\gamma P_{1}\hat{a}_{0}'(\omega,z)e^{i2\gamma P_{1}z}=\\
 &  & 2i\gamma\times\sqrt{P_{1}}\sqrt{\zeta_{2}}\times\hat{a}_{0}'^\dagger(\omega_{p1}+\omega_{p2}-\omega,z)\hat{a}_{0}'(\omega_{p2},z)e^{-i\left(\Delta k-\gamma P_{1}\right)z}+2i\gamma\times P_{1}\hat{a}_{0}'(\omega,z)e^{i2\gamma P_{1}z}\nonumber
\end{eqnarray}
so
\begin{eqnarray}
  \frac{\partial \hat{a}_{0}'(\omega,z)}{\partial z}= 2i\gamma\times\sqrt{P_{1}}\sqrt{\zeta_{2}}\times\hat{a}_{0}'^\dagger(\omega_{p1}+\omega_{p2}-\omega,z)\hat{a}_{0}'(\omega_{p2},z)e^{-i\left(\Delta k+\gamma P_{1}\right)z} \label{eq:daw/dz}.
\end{eqnarray}
For the weak pump it gives
\begin{eqnarray}
 &  & \frac{\partial \hat{a}_{0}'(\omega_{p2},z)}{\partial z}e^{i2\gamma P_{1}z}+i2\gamma P_{1}\hat{a}_{0}'(\omega_{p2},z)e^{i2\gamma P_{1}z}=\\
 &  & 2i\gamma\times P_{1}\hat{a}_{0}'(\omega_{p2},z)e^{i2\gamma P_{1}z}+2i\gamma\times\sqrt{P_{1}}\sqrt{\zeta_{2}}\times\frac{T}{2\pi}\times\int d\omega\sqrt{\frac{\omega\left(\omega_{p1}+\omega_{p2}-\omega\right)}{\omega_{p1}\omega_{p2}}}\hat{a}'_{0}(\omega,z)\hat{a}'_{0}(\omega_{p1}+\omega_{p2}-\omega,z)e^{i\left(\Delta k+\gamma P_{1}\right)z}\nonumber,
\end{eqnarray}
or

\begin{eqnarray}
  \frac{\partial \hat{a}_{0}'(\omega_{p2},z)}{\partial z}=
 2i\gamma\times\sqrt{P_{1}}\sqrt{\zeta_{2}}\times\frac{T}{2\pi}\int d\omega\sqrt{\frac{\omega\left(\omega_{p1}+\omega_{p2}-\omega\right)}{\omega_{p1}\omega_{p2}}}\hat{a}_{0}'(\omega,z)\hat{a}_{0}'(\omega_{p1}+\omega_{p2}-\omega,z)e^{i\left(\Delta k+\gamma P_{1}\right)z}\label{eq:dawp2/dz} .
\end{eqnarray}
\end{widetext}
The total phase mismatch is now $K=\Delta k+\gamma P_{1}$. The modes' evolution can be solved by performing a Baker-Hausdorff expansion. If the $z$ evolution of the momentum operator
$\hat{G}(z)$ is slow enough to be considered as $z$ independent, which is the case in a low gain interaction, $\int_{0}^{L}\hat{G_{nl}}(z)dz\simeq\hat{G_{nl}}L$ and $\hat{a}_{0}(\omega,L)=e^{-\frac{i}{\hbar}\hat{G_{nl}}L}\hat{a}_{0}(\omega,0)e^{+\frac{i}{\hbar}\hat{G_{nl}}L}$,
which gives

\begin{eqnarray}
 & \hat{a}_{0}(\omega,L) & =\hat{a}_{0}(\omega,0)+\left[\hat{a}_{0}(\omega,0),\frac{i}{\hbar}\hat{G_{nl}}L\right]+\\
 & & \frac{1}{2!}\left[\left[\hat{a}_{0}(\omega,0),\frac{i}{\hbar}\hat{G_{nl}}L\right],\frac{i}{\hbar}\hat{G_{nl}}L\right]+\nonumber \\
 &  & \frac{1}{3!}\left[\left[\left[\hat{a}_{0}(\omega,0),\frac{i}{\hbar}\hat{G_{nl}}L\right],\frac{i}{\hbar}\hat{G_{nl}}L\right],\frac{i}{\hbar}\hat{G_{nl}}L\right]+...\nonumber
\end{eqnarray}
or, given that $\frac{i}{\hbar}\left[\hat{a}_{0}(\omega,0),\hat{G}_{nl}(L)\right]=\left(\frac{\partial\hat{a}_{0}(\omega,z)}{\partial z}\right)_{z=0}$,

\begin{eqnarray}
&&\hat{a}_{0}(\omega,L)=\hat{a}_{0}(\omega,0)+\left(\frac{\partial\hat{a}_{0}(\omega,z)}{\partial z}\right)_{z=0}L+\label{eq:da/dz taylor}\\
&& \left(\frac{\partial^{2}\hat{a}_{0}(\omega,z)}{\partial z^{2}}\right)_{z=0}\frac{L^{2}}{2!}+\left(\frac{\partial^{3}\hat{a}_{0}(\omega,z)}{\partial z^{3}}\right)_{z=0}\frac{L^{3}}{3!}+...~.\nonumber
\end{eqnarray}
This development (\ref{eq:da/dz taylor}), equivalent to a Taylor expansion for the operators, gives the creation and annihilation operators at the output of the medium for the generated modes and the weak pump, solutions of Eqs.~(\ref{eq:daw/dz})
and (\ref{eq:dawp2/dz}) respectively. Let's solve it for the generated modes.

Eq.~(\ref{eq:daw/dz}) gives for the higher order derivatives 
\begin{small}
\begin{eqnarray}
 &  & \frac{\partial^{2}\hat{a}_{0}'(\omega,z)}{\partial z^{2}}=\label{eq:daw/dz-1}\\
 &  & -iK\times2i\gamma\sqrt{P_{1}}\sqrt{\zeta_{2}}\hat{a}_{0}'^\dagger(\omega_{p1}+\omega_{p2}-\omega,z)\hat{a}_{0}'(\omega_{p2},z)e^{-iKz}+\nonumber \\
 &  & 2i\gamma\sqrt{P_{1}}\sqrt{\zeta_{2}}\times\frac{\partial\hat{a}_{0}'^\dagger(\omega_{p1}+\omega_{p2}-\omega,z)}{\partial z}\hat{a}_{0}'(\omega_{p2},z)e^{-iKz}+\nonumber \\
 &  & 2i\gamma\sqrt{P_{1}}\sqrt{\zeta_{2}}\times\hat{a}_{0}'^\dagger(\omega_{p1}+\omega_{p2}-\omega,z)\frac{\partial\hat{a}_{0}'(\omega_{p2},z)}{\partial z}e^{-iKz}\nonumber 
\end{eqnarray}
\end{small}

so
\begin{small}
\begin{eqnarray}
 &  & \frac{\partial^{2}\hat{a}_{0}'(\omega,z)}{\partial z^{2}}=\label{eq:daw/dz-1-1}\\
 &  & -iK\times2i\gamma\sqrt{P_{1}}\sqrt{\zeta_{2}}~\hat{a}_{0}'^\dagger(\omega_{p1}+\omega_{p2}-\omega,z)\hat{a}_{0}'(\omega_{p2},z)e^{-iKz}+\nonumber \\
 &  & \left(2i\gamma\sqrt{P_{1}}\sqrt{\zeta_{2}}\right)^{2}\times\hat{a}_{0}'^\dagger(\omega,z)\hat{a}_{0}'(\omega_{p2},z)\hat{a}_{0}'(\omega_{p2},z)e^{-iKz}e^{iKz}+\nonumber \\
 &  & \left(2i\gamma\sqrt{P_{1}}\sqrt{\zeta_{2}}\right)^{2}\times\frac{T}{2\pi}\int d\omega\sqrt{\frac{\omega\left(\omega_{p1}+\omega_{p2}-\omega\right)}{\omega_{p1}\omega_{p2}}}\times\nonumber \\
 &  & \hat{a}_{0}'^\dagger(\omega_{p1}+\omega_{p2}-\omega,z)\hat{a}_{0}'(\omega,z)\hat{a}_{0}'(\omega_{p1}+\omega_{p2}-\omega,z)e^{-iKz}e^{-iKz}\nonumber.
\end{eqnarray}
\end{small}
The quantity $\gamma\times\sqrt{P_{1}}\sqrt{\frac{T}{2\pi}\zeta_{2}}$ being very
small (it is the square root number of photons generated
in 1 meter of medium within a frequency range of $\delta\omega$, see the link to the efficieny $\eta$ in Eq. (\ref{eq:efficiency})), we will limit our development to the first order in $\gamma\times\sqrt{P_{1}}\sqrt{\frac{T}{2\pi}\zeta_{2}}$. This approximation is physically equivalent to neglecting all the phenomena involving more than a single pair creation, in particular here the recombination of a created pair back into pump photons, and the Rabi oscillations that can then occur between signal/idler and pump photon states. Then we have
\begin{widetext} 
\begin{eqnarray}
\frac{\partial^{2}\hat{a}_{0}'(\omega,z)}{\partial z^{2}}= -iK\times2i\gamma\sqrt{P_{1}}\sqrt{\zeta_{2}}\times\hat{a}_{0}'^\dagger(\omega_{p1}+\omega_{p2}-\omega,z)\hat{a}_{0}'(\omega_{p2},z)e^{-iKz}.
\end{eqnarray}
By doing the same approximation for the third order we get 
\begin{eqnarray}
 \frac{\partial^{3}\hat{a}_{0}'(\omega,z)}{\partial z^{3}}=\left(-iK\right)^{2}2i\gamma\sqrt{P_{1}}\sqrt{\zeta_{2}}\times\hat{a}_{0}'^\dagger(\omega_{p1}+\omega_{p2}-\omega,z)\hat{a}_{0}'(\omega_{p2},z)e^{-iKz},
\end{eqnarray}
and by an obvious recurrence 

\begin{eqnarray}
 \frac{\partial^{n}\hat{a}_{0}'(\omega,z)}{\partial z^{n}}=\left(-iK\right)^{n-1}2i\gamma\sqrt{P_{1}}\sqrt{\zeta_{2}}\times\hat{a}_{0}'(\omega_{p1}+\omega_{p2}-\omega,z)\hat{a}_{0}^{'+}(\omega_{p2},z)e^{-iKz}.
\end{eqnarray}

Then Eq.~(\ref{eq:da/dz taylor}) gives 

\begin{eqnarray}
\hat{a}_{0}'(\omega,L)   = \hat{a}_{0}'(\omega,0)+2i\gamma\sqrt{P_{1}}\sqrt{\zeta_{2}}\hat{a}_{0}'^\dagger(\omega_{p1}+\omega_{p2}-\omega,0)\hat{a}_{0}'(\omega_{p2},0)e^{-iKL}\sum_{n=1}^{+\infty}\frac{L^{n}}{n!}(-iK)^{n-1}\nonumber.
\end{eqnarray}

The power series can be simplified to

\begin{eqnarray} 
\sum_{n=1}^{+\infty}\frac{L^{n}}{n!}\left(-iK\right)^{n-1}= L\sum_{n=1}^{+\infty}\frac{\left(-iKL\right)}{n!}^{n-1}=
  L\frac{e^{-iKL}-1}{-iKL}= e^{-\frac{iKL}{2}}L\mbox{sinc}\left(\frac{KL}{2}\right),
\end{eqnarray}
so
\begin{eqnarray}
\hat{a}_{0}'(\omega,L)  =  \hat{a}_{0}'(\omega,0)+2i\gamma\times\sqrt{P_{1}}\sqrt{\zeta_{2}}e^{-\frac{iKL}{2}}L\times\mbox{sinc}\left(\frac{KL}{2}\right)\times\hat{a}_{0}'^\dagger(\omega_{p1}+\omega_{p2}-\omega,0)\hat{a}_{0}'(\omega_{p2},0).
\nonumber\end{eqnarray}
Then with $\hat{a}_{0}(\omega,z)=\hat{a}_{0}'(\omega,z)e^{i2\gamma P_{1}z}$,
\begin{eqnarray}
\hat{a}_{0}(\omega,L)e^{-i2\gamma P_{1}L} & = & \hat{a}_{0}(\omega,0)+2i\gamma\sqrt{P_{1}}\sqrt{\zeta_{2}}e^{-\frac{iKL}{2}}L\times\mbox{sinc}\left(\frac{KL}{2}\right)\hat{a}_{0}{}^\dagger(\omega_{p1}+\omega_{p2}-\omega,0)\hat{a}_{0}(\omega_{p2},0).
\nonumber\end{eqnarray}
\end{widetext}
\section{Validity of low gain approximation}\label{higher_order_in_gain}
In Appendix~\ref{app:BCH} we developed the creation and annihilation operator evolution to the first order in the gain. If $\eta$, as defined in Eq.~(\ref{eq:efficiency}), is much less than 1, it is the efficiency of conversion of a single photon into a pair. Searching for the precision of this result, we have to go to higher orders in gain in the development of Appendix~\ref{app:BCH}. Going to the second order in $\eta$ allows the possibility of having a pair converted back into a single photon, which occurs with probability $\eta^2$, but the forward single photon conversion efficiency is unchanged. Going to the third order in $\eta$ then lowers the single photon conversion efficiency to $\eta-2\eta^2+\eta^3$. This result allows us to define the error due to taking the gain to first order as $2\eta^2$.

In principle, $\eta$ could have an arbitrarily large value
by increasing the strong pump power. When $\eta$ approaches and 
goes beyond 1, it cannot be defined as a probability of conversion, and we must solve exactly the operator
evolution. As developed in a classical setting in Ref.~\cite{Chen1989}, elliptic functions are expected for the beams' intensity evolution, giving Rabi-like oscillations between the pump photon and the signal/idler pair
state. As noted in Ref.~\cite{Langford2011Efficien}, the higher the $\eta$, the more oscillations will occur in the fiber, but this is again a theoretical scheme taking only nondegenerate FWM into account. Parasitic phenomena, in particular self-phase modulation, will also become ultra high in this regime.


\begin{thebibliography}{27}%
\makeatletter
\providecommand \@ifxundefined [1]{%
 \@ifx{#1\undefined}
}%
\providecommand \@ifnum [1]{%
 \ifnum #1\expandafter \@firstoftwo
 \else \expandafter \@secondoftwo
 \fi
}%
\providecommand \@ifx [1]{%
 \ifx #1\expandafter \@firstoftwo
 \else \expandafter \@secondoftwo
 \fi
}%
\providecommand \natexlab [1]{#1}%
\providecommand \enquote  [1]{``#1''}%
\providecommand \bibnamefont  [1]{#1}%
\providecommand \bibfnamefont [1]{#1}%
\providecommand \citenamefont [1]{#1}%
\providecommand \href@noop [0]{\@secondoftwo}%
\providecommand \href [0]{\begingroup \@sanitize@url \@href}%
\providecommand \@href[1]{\@@startlink{#1}\@@href}%
\providecommand \@@href[1]{\endgroup#1\@@endlink}%
\providecommand \@sanitize@url [0]{\catcode `\\12\catcode `\$12\catcode
  `\&12\catcode `\#12\catcode `\^12\catcode `\_12\catcode `\%12\relax}%
\providecommand \@@startlink[1]{}%
\providecommand \@@endlink[0]{}%
\providecommand \url  [0]{\begingroup\@sanitize@url \@url }%
\providecommand \@url [1]{\endgroup\@href {#1}{\urlprefix }}%
\providecommand \urlprefix  [0]{URL }%
\providecommand \Eprint [0]{\href }%
\providecommand \doibase [0]{http://dx.doi.org/}%
\providecommand \selectlanguage [0]{\@gobble}%
\providecommand \bibinfo  [0]{\@secondoftwo}%
\providecommand \bibfield  [0]{\@secondoftwo}%
\providecommand \translation [1]{[#1]}%
\providecommand \BibitemOpen [0]{}%
\providecommand \bibitemStop [0]{}%
\providecommand \bibitemNoStop [0]{.\EOS\space}%
\providecommand \EOS [0]{\spacefactor3000\relax}%
\providecommand \BibitemShut  [1]{\csname bibitem#1\endcsname}%
\let\auto@bib@innerbib\@empty
%</preamble>
\bibitem [{\citenamefont {Burnham}\ and\ \citenamefont
  {Weinberg}(1970)}]{Burnham_SPDC_70}%
  \BibitemOpen
  \bibfield  {author} {\bibinfo {author} {\bibfnamefont {D.~C.}\ \bibnamefont
  {Burnham}}\ and\ \bibinfo {author} {\bibfnamefont {D.~L.}\ \bibnamefont
  {Weinberg}},\ }\href@noop {} {\bibfield  {journal} {\bibinfo  {journal}
  {Phys. Rev. Lett.}\ }\textbf {\bibinfo {volume} {25}},\ \bibinfo {pages} {84}
  (\bibinfo {year} {1970})}\BibitemShut {NoStop}%
\bibitem [{\citenamefont {Fiorentino}\ \emph {et~al.}(2002)\citenamefont
  {Fiorentino}, \citenamefont {Voss}, \citenamefont {Sharping},\ and\
  \citenamefont {Kumar}}]{Fiorentino2002-sp-FWM}%
  \BibitemOpen
  \bibfield  {author} {\bibinfo {author} {\bibfnamefont {M.}~\bibnamefont
  {Fiorentino}}, \bibinfo {author} {\bibfnamefont {P.}~\bibnamefont {Voss}},
  \bibinfo {author} {\bibfnamefont {J.}~\bibnamefont {Sharping}}, \ and\
  \bibinfo {author} {\bibfnamefont {P.}~\bibnamefont {Kumar}},\ }\href
  {\doibase 10.1109/LPT.2002.1012406} {\bibfield  {journal} {\bibinfo
  {journal} {IEEE Photonics Technology Letters}\ }\textbf {\bibinfo {volume}
  {14}},\ \bibinfo {pages} {983} (\bibinfo {year} {2002})}\BibitemShut
  {NoStop}%
\bibitem [{\citenamefont {Guerreiro}\ \emph {et~al.}(2013)\citenamefont
  {Guerreiro}, \citenamefont {Pomarico}, \citenamefont {Sanguinetti},
  \citenamefont {Sangouard}, \citenamefont {Pelc}, \citenamefont {Langrock},
  \citenamefont {Fejer}, \citenamefont {Zbinden}, \citenamefont {Thew},\ and\
  \citenamefont {Gisin}}]{Guerreiro2013Interact}%
  \BibitemOpen
  \bibfield  {author} {\bibinfo {author} {\bibfnamefont {T.}~\bibnamefont
  {Guerreiro}}, \bibinfo {author} {\bibfnamefont {E.}~\bibnamefont {Pomarico}},
  \bibinfo {author} {\bibfnamefont {B.}~\bibnamefont {Sanguinetti}}, \bibinfo
  {author} {\bibfnamefont {N.}~\bibnamefont {Sangouard}}, \bibinfo {author}
  {\bibfnamefont {J.~S.}\ \bibnamefont {Pelc}}, \bibinfo {author}
  {\bibfnamefont {C.}~\bibnamefont {Langrock}}, \bibinfo {author}
  {\bibfnamefont {M.~M.}\ \bibnamefont {Fejer}}, \bibinfo {author}
  {\bibfnamefont {H.}~\bibnamefont {Zbinden}}, \bibinfo {author} {\bibfnamefont
  {R.~T.}\ \bibnamefont {Thew}}, \ and\ \bibinfo {author} {\bibfnamefont
  {N.}~\bibnamefont {Gisin}},\ }\href {http://dx.doi.org/10.1038/ncomms3324}
  {\bibfield  {journal} {\bibinfo  {journal} {Nat. Commun.}\ }\textbf {\bibinfo
  {volume} {4}},\ \bibinfo {pages} {2324} (\bibinfo {year} {2013})}\BibitemShut
  {NoStop}%
\bibitem [{\citenamefont {{Hamel}}\ \emph {et~al.}(2014)\citenamefont
  {{Hamel}}, \citenamefont {{Shalm}}, \citenamefont {{H{\"u}bel}},
  \citenamefont {{Miller}}, \citenamefont {{Marsili}}, \citenamefont {{Verma}},
  \citenamefont {{Mirin}}, \citenamefont {{Nam}}, \citenamefont {{Resch}},\
  and\ \citenamefont {{Jennewein}}}]{2014arXiv1404.7131H}%
  \BibitemOpen
  \bibfield  {author} {\bibinfo {author} {\bibfnamefont {D.~R.}\ \bibnamefont
  {{Hamel}}}, \bibinfo {author} {\bibfnamefont {L.~K.}\ \bibnamefont
  {{Shalm}}}, \bibinfo {author} {\bibfnamefont {H.}~\bibnamefont
  {{H{\"u}bel}}}, \bibinfo {author} {\bibfnamefont {A.~J.}\ \bibnamefont
  {{Miller}}}, \bibinfo {author} {\bibfnamefont {F.}~\bibnamefont {{Marsili}}},
  \bibinfo {author} {\bibfnamefont {V.~B.}\ \bibnamefont {{Verma}}}, \bibinfo
  {author} {\bibfnamefont {R.~P.}\ \bibnamefont {{Mirin}}}, \bibinfo {author}
  {\bibfnamefont {S.~W.}\ \bibnamefont {{Nam}}}, \bibinfo {author}
  {\bibfnamefont {K.~J.}\ \bibnamefont {{Resch}}}, \ and\ \bibinfo {author}
  {\bibfnamefont {T.}~\bibnamefont {{Jennewein}}},\ }\href@noop {} {\bibfield
  {journal} {\bibinfo  {journal} {ArXiv e-prints}\ } (\bibinfo {year}
  {2014})},\ \Eprint {http://arxiv.org/abs/1404.7131} {arXiv:1404.7131
  [quant-ph]} \BibitemShut {NoStop}%
\bibitem [{\citenamefont {Miwa}\ \emph {et~al.}(2014)\citenamefont {Miwa},
  \citenamefont {Yoshikawa}, \citenamefont {Iwata}, \citenamefont {Endo},
  \citenamefont {Marek}, \citenamefont {Filip}, \citenamefont {van Loock},\
  and\ \citenamefont {Furusawa}}]{PhysRevLett.113.013601}%
  \BibitemOpen
  \bibfield  {author} {\bibinfo {author} {\bibfnamefont {Y.}~\bibnamefont
  {Miwa}}, \bibinfo {author} {\bibfnamefont {J.-i.}\ \bibnamefont {Yoshikawa}},
  \bibinfo {author} {\bibfnamefont {N.}~\bibnamefont {Iwata}}, \bibinfo
  {author} {\bibfnamefont {M.}~\bibnamefont {Endo}}, \bibinfo {author}
  {\bibfnamefont {P.}~\bibnamefont {Marek}}, \bibinfo {author} {\bibfnamefont
  {R.}~\bibnamefont {Filip}}, \bibinfo {author} {\bibfnamefont
  {P.}~\bibnamefont {van Loock}}, \ and\ \bibinfo {author} {\bibfnamefont
  {A.}~\bibnamefont {Furusawa}},\ }\href {\doibase
  10.1103/PhysRevLett.113.013601} {\bibfield  {journal} {\bibinfo  {journal}
  {Phys. Rev. Lett.}\ }\textbf {\bibinfo {volume} {113}},\ \bibinfo {pages}
  {013601} (\bibinfo {year} {2014})}\BibitemShut {NoStop}%
\bibitem [{\citenamefont {Huebel}\ \emph {et~al.}(2010)\citenamefont {Huebel},
  \citenamefont {Hamel}, \citenamefont {Fedrizzi}, \citenamefont {Ramelow},
  \citenamefont {Resch},\ and\ \citenamefont {Jennewein}}]{Hubel2010Direct-g}%
  \BibitemOpen
  \bibfield  {author} {\bibinfo {author} {\bibfnamefont {H.}~\bibnamefont
  {Huebel}}, \bibinfo {author} {\bibfnamefont {D.~R.}\ \bibnamefont {Hamel}},
  \bibinfo {author} {\bibfnamefont {A.}~\bibnamefont {Fedrizzi}}, \bibinfo
  {author} {\bibfnamefont {S.}~\bibnamefont {Ramelow}}, \bibinfo {author}
  {\bibfnamefont {K.~J.}\ \bibnamefont {Resch}}, \ and\ \bibinfo {author}
  {\bibfnamefont {T.}~\bibnamefont {Jennewein}},\ }\href@noop {} {\bibfield
  {journal} {\bibinfo  {journal} {Nature}\ }\textbf {\bibinfo {volume} {466}},\
  \bibinfo {pages} {601} (\bibinfo {year} {2010})}\BibitemShut {NoStop}%
\bibitem [{\citenamefont {Shalm}\ \emph {et~al.}(2013)\citenamefont {Shalm},
  \citenamefont {Hamel}, \citenamefont {Yan}, \citenamefont {Simon},
  \citenamefont {Resch},\ and\ \citenamefont {Jennewein}}]{Shalm2013Three-ph}%
  \BibitemOpen
  \bibfield  {author} {\bibinfo {author} {\bibfnamefont {L.~K.}\ \bibnamefont
  {Shalm}}, \bibinfo {author} {\bibfnamefont {D.~R.}\ \bibnamefont {Hamel}},
  \bibinfo {author} {\bibfnamefont {Z.}~\bibnamefont {Yan}}, \bibinfo {author}
  {\bibfnamefont {C.}~\bibnamefont {Simon}}, \bibinfo {author} {\bibfnamefont
  {K.~J.}\ \bibnamefont {Resch}}, \ and\ \bibinfo {author} {\bibfnamefont
  {T.}~\bibnamefont {Jennewein}},\ }\href@noop {} {\bibfield  {journal}
  {\bibinfo  {journal} {Nat. Phys.}\ }\textbf {\bibinfo {volume} {9}},\
  \bibinfo {pages} {19} (\bibinfo {year} {2013})}\BibitemShut {NoStop}%
\bibitem [{\citenamefont {Browne}\ and\ \citenamefont
  {Rudolph}(2005)}]{Browne_Rudolph2005}%
  \BibitemOpen
  \bibfield  {author} {\bibinfo {author} {\bibfnamefont {D.~E.}\ \bibnamefont
  {Browne}}\ and\ \bibinfo {author} {\bibfnamefont {T.}~\bibnamefont
  {Rudolph}},\ }\href@noop {} {\bibfield  {journal} {\bibinfo  {journal} {Phys.
  Rev. Lett.}\ }\textbf {\bibinfo {volume} {95}},\ \bibinfo {pages} {010501}
  (\bibinfo {year} {2005})}\BibitemShut {NoStop}%
\bibitem [{\citenamefont {Hillery}\ \emph {et~al.}(1999)\citenamefont
  {Hillery}, \citenamefont {Buzek},\ and\ \citenamefont
  {Berthiaume}}]{Hillery_1999}%
  \BibitemOpen
  \bibfield  {author} {\bibinfo {author} {\bibfnamefont {M.}~\bibnamefont
  {Hillery}}, \bibinfo {author} {\bibfnamefont {V.}~\bibnamefont {Buzek}}, \
  and\ \bibinfo {author} {\bibfnamefont {A.}~\bibnamefont {Berthiaume}},\
  }\href {\doibase 10.1103/PhysRevA.59.1829} {\bibfield  {journal} {\bibinfo
  {journal} {Phys. Rev. A}\ }\textbf {\bibinfo {volume} {59}},\ \bibinfo
  {pages} {1829} (\bibinfo {year} {1999})}\BibitemShut {NoStop}%
\bibitem [{\citenamefont {Banaszek}\ and\ \citenamefont
  {Knight}(1997)}]{banaszek1997qit}%
  \BibitemOpen
  \bibfield  {author} {\bibinfo {author} {\bibfnamefont {K.}~\bibnamefont
  {Banaszek}}\ and\ \bibinfo {author} {\bibfnamefont {P.~L.}\ \bibnamefont
  {Knight}},\ }\href@noop {} {\bibfield  {journal} {\bibinfo  {journal} {Phys.
  Rev. A}\ }\textbf {\bibinfo {volume} {55}},\ \bibinfo {pages} {2368}
  (\bibinfo {year} {1997})}\BibitemShut {NoStop}%
\bibitem [{\citenamefont {Greenberger}\ \emph {et~al.}(1990)\citenamefont
  {Greenberger}, \citenamefont {Horne}, \citenamefont {Shimony},\ and\
  \citenamefont {Zeilinger}}]{greenberger1990bst}%
  \BibitemOpen
  \bibfield  {author} {\bibinfo {author} {\bibfnamefont {D.~M.}\ \bibnamefont
  {Greenberger}}, \bibinfo {author} {\bibfnamefont {M.~A.}\ \bibnamefont
  {Horne}}, \bibinfo {author} {\bibfnamefont {A.}~\bibnamefont {Shimony}}, \
  and\ \bibinfo {author} {\bibfnamefont {A.}~\bibnamefont {Zeilinger}},\
  }\href@noop {} {\bibfield  {journal} {\bibinfo  {journal} {Am. J. Phys.}\
  }\textbf {\bibinfo {volume} {58}},\ \bibinfo {pages} {1131} (\bibinfo {year}
  {1990})}\BibitemShut {NoStop}%
\bibitem [{\citenamefont {Cabello}\ and\ \citenamefont
  {Sciarrino}(2012)}]{PhysRevX.2.021010}%
  \BibitemOpen
  \bibfield  {author} {\bibinfo {author} {\bibfnamefont {A.}~\bibnamefont
  {Cabello}}\ and\ \bibinfo {author} {\bibfnamefont {F.}~\bibnamefont
  {Sciarrino}},\ }\href {\doibase 10.1103/PhysRevX.2.021010} {\bibfield
  {journal} {\bibinfo  {journal} {Phys. Rev. X}\ }\textbf {\bibinfo {volume}
  {2}},\ \bibinfo {pages} {21010} (\bibinfo {year} {2012})}\BibitemShut
  {NoStop}%
\bibitem [{\citenamefont {Gisin}\ \emph {et~al.}(2010)\citenamefont {Gisin},
  \citenamefont {Pironio},\ and\ \citenamefont
  {Sangouard}}]{PhysRevLett.105.070501}%
  \BibitemOpen
  \bibfield  {author} {\bibinfo {author} {\bibfnamefont {N.}~\bibnamefont
  {Gisin}}, \bibinfo {author} {\bibfnamefont {S.}~\bibnamefont {Pironio}}, \
  and\ \bibinfo {author} {\bibfnamefont {N.}~\bibnamefont {Sangouard}},\ }\href
  {\doibase 10.1103/PhysRevLett.105.070501} {\bibfield  {journal} {\bibinfo
  {journal} {Phys. Rev. Lett.}\ }\textbf {\bibinfo {volume} {105}},\ \bibinfo
  {pages} {070501} (\bibinfo {year} {2010})}\BibitemShut {NoStop}%
\bibitem [{\citenamefont {Langford}\ \emph {et~al.}(2011)\citenamefont
  {Langford}, \citenamefont {Ramelow}, \citenamefont {Prevedel}, \citenamefont
  {Munro}, \citenamefont {Milburn},\ and\ \citenamefont
  {Zeilinger}}]{Langford2011Efficien}%
  \BibitemOpen
  \bibfield  {author} {\bibinfo {author} {\bibfnamefont {N.~K.}\ \bibnamefont
  {Langford}}, \bibinfo {author} {\bibfnamefont {S.}~\bibnamefont {Ramelow}},
  \bibinfo {author} {\bibfnamefont {R.}~\bibnamefont {Prevedel}}, \bibinfo
  {author} {\bibfnamefont {W.~J.}\ \bibnamefont {Munro}}, \bibinfo {author}
  {\bibfnamefont {G.~J.}\ \bibnamefont {Milburn}}, \ and\ \bibinfo {author}
  {\bibfnamefont {A.}~\bibnamefont {Zeilinger}},\ }\href@noop {} {\bibfield
  {journal} {\bibinfo  {journal} {Nature}\ }\textbf {\bibinfo {volume} {478}},\
  \bibinfo {pages} {360} (\bibinfo {year} {2011})}\BibitemShut {NoStop}%
\bibitem [{\citenamefont {Tanzilli}\ \emph {et~al.}(2001)\citenamefont
  {Tanzilli}, \citenamefont {{De Riedmatten}}, \citenamefont {Tittel},
  \citenamefont {Zbinden}, \citenamefont {Baldi}, \citenamefont {{De Micheli}},
  \citenamefont {Ostrowsky},\ and\ \citenamefont {Gisin}}]{Tanzilli4Highly-e}%
  \BibitemOpen
  \bibfield  {author} {\bibinfo {author} {\bibfnamefont {S.}~\bibnamefont
  {Tanzilli}}, \bibinfo {author} {\bibfnamefont {H.}~\bibnamefont {{De
  Riedmatten}}}, \bibinfo {author} {\bibfnamefont {W.}~\bibnamefont {Tittel}},
  \bibinfo {author} {\bibfnamefont {H.}~\bibnamefont {Zbinden}}, \bibinfo
  {author} {\bibfnamefont {P.}~\bibnamefont {Baldi}}, \bibinfo {author}
  {\bibfnamefont {M.}~\bibnamefont {{De Micheli}}}, \bibinfo {author}
  {\bibfnamefont {D.~B.}\ \bibnamefont {Ostrowsky}}, \ and\ \bibinfo {author}
  {\bibfnamefont {N.}~\bibnamefont {Gisin}},\ }\href {\doibase
  10.1049/el:20010009} {\bibfield  {journal} {\bibinfo  {journal} {Electron.
  Lett.}\ }\textbf {\bibinfo {volume} {37}},\ \bibinfo {pages} {26} (\bibinfo
  {year} {2001})}\BibitemShut {NoStop}%
\bibitem [{\citenamefont {Brainis}(2009)}]{Brainis2009}%
  \BibitemOpen
  \bibfield  {author} {\bibinfo {author} {\bibfnamefont {E.}~\bibnamefont
  {Brainis}},\ }\href {\doibase 10.1103/PhysRevA.79.023840} {\bibfield
  {journal} {\bibinfo  {journal} {Physical Review A}\ }\textbf {\bibinfo
  {volume} {79}},\ \bibinfo {pages} {023840} (\bibinfo {year}
  {2009})}\BibitemShut {NoStop}%
\bibitem [{\citenamefont {Lin}\ \emph {et~al.}(2007)\citenamefont {Lin},
  \citenamefont {Yaman},\ and\ \citenamefont {Agrawal}}]{Agrawal2007}%
  \BibitemOpen
  \bibfield  {author} {\bibinfo {author} {\bibfnamefont {Q.}~\bibnamefont
  {Lin}}, \bibinfo {author} {\bibfnamefont {F.}~\bibnamefont {Yaman}}, \ and\
  \bibinfo {author} {\bibfnamefont {G.P.}~\bibnamefont {Agrawal}},\ }\href
  {\doibase 10.1103/PhysRevA.75.023803} {\bibfield  {journal} {\bibinfo
  {journal} {Physical Review A}\ }\textbf {\bibinfo {volume} {75}},\ \bibinfo
  {pages} {023803} (\bibinfo {year} {2007})}\BibitemShut {NoStop}%
\bibitem [{\citenamefont {Huttner}\ \emph {et~al.}(1990)\citenamefont
  {Huttner}, \citenamefont {Serulnik},\ and\ \citenamefont
  {Ben-Aryeh}}]{Huttner1990}%
  \BibitemOpen
  \bibfield  {author} {\bibinfo {author} {\bibfnamefont {B.}~\bibnamefont
  {Huttner}}, \bibinfo {author} {\bibfnamefont {S.}~\bibnamefont {Serulnik}}, \
  and\ \bibinfo {author} {\bibfnamefont {Y.}~\bibnamefont {Ben-Aryeh}},\
  }\href@noop {} {\bibfield  {journal} {\bibinfo  {journal} {Phys. Rev. A}\
  }\textbf {\bibinfo {volume} {42}},\ \bibinfo {pages} {5594} (\bibinfo {year}
  {1990})}\BibitemShut {NoStop}%
\bibitem [{\citenamefont {Blow}\ \emph {et~al.}(1990)\citenamefont {Blow},
  \citenamefont {Loudon}, \citenamefont {Phoenix},\ and\ \citenamefont
  {Shepherd}}]{Loudon1990}%
  \BibitemOpen
  \bibfield  {author} {\bibinfo {author} {\bibfnamefont {K.~J.}\ \bibnamefont
  {Blow}}, \bibinfo {author} {\bibfnamefont {R.}~\bibnamefont {Loudon}},
  \bibinfo {author} {\bibfnamefont {S.~J.~D.}\ \bibnamefont {Phoenix}}, \ and\
  \bibinfo {author} {\bibfnamefont {T.~J.}\ \bibnamefont {Shepherd}},\ }\href
  {\doibase 10.1103/PhysRevA.42.4102} {\bibfield  {journal} {\bibinfo
  {journal} {Phys. Rev. A}\ }\textbf {\bibinfo {volume} {42}},\ \bibinfo
  {pages} {4102} (\bibinfo {year} {1990})}\BibitemShut {NoStop}%
\bibitem [{\citenamefont {Agrawal}(2006)}]{Agrawal-NL-Fibre-Optics}%
  \BibitemOpen
  \bibfield  {author} {\bibinfo {author} {\bibfnamefont {G.}~\bibnamefont
  {Agrawal}},\ }\href
  {http://www.amazon.com/exec/obidos/redirect?tag=citeulike07-20\&path=ASIN/0120451433}
  {\emph {\bibinfo {title} {{Nonlinear Fiber Optics}}}},\ \bibinfo {edition}
  {4th}\ ed.\ (\bibinfo  {publisher} {Academic Press},\ \bibinfo {year}
  {2006})\BibitemShut {NoStop}%
\bibitem [{Note1()}]{Note1}%
  \BibitemOpen
  \bibinfo {note} {Material absorption can be included here and in all the
  following evolutions by taking $z$ as the effective position in the fiber;
  $z={\begingroup 1-e^{-\alpha z^\prime }\endgroup \over \alpha }$ for
  absorption coefficient $\alpha $ and true position $z^\prime $.}\BibitemShut
  {Stop}%
\bibitem [{\citenamefont {Smith}\ \emph {et~al.}(2009)\citenamefont {Smith},
  \citenamefont {Mahou}, \citenamefont {Cohen}, \citenamefont {Lundeen},\ and\
  \citenamefont {Walmsley}}]{Smith:09}%
  \BibitemOpen
  \bibfield  {author} {\bibinfo {author} {\bibfnamefont {B.~J.}\ \bibnamefont
  {Smith}}, \bibinfo {author} {\bibfnamefont {P.}~\bibnamefont {Mahou}},
  \bibinfo {author} {\bibfnamefont {O.}~\bibnamefont {Cohen}}, \bibinfo
  {author} {\bibfnamefont {J.~S.}\ \bibnamefont {Lundeen}}, \ and\ \bibinfo
  {author} {\bibfnamefont {I.~A.}\ \bibnamefont {Walmsley}},\ }\href {\doibase
  10.1364/OE.17.023589} {\bibfield  {journal} {\bibinfo  {journal} {Opt.
  Express}\ }\textbf {\bibinfo {volume} {17}},\ \bibinfo {pages} {23589}
  (\bibinfo {year} {2009})}\BibitemShut {NoStop}%
\bibitem [{\citenamefont {Rarity}\ \emph {et~al.}(2005)\citenamefont {Rarity},
  \citenamefont {Fulconis}, \citenamefont {Duligall}, \citenamefont
  {Wadsworth},\ and\ \citenamefont {Russell}}]{Rarity:05}%
  \BibitemOpen
  \bibfield  {author} {\bibinfo {author} {\bibfnamefont {J.}~\bibnamefont
  {Rarity}}, \bibinfo {author} {\bibfnamefont {J.}~\bibnamefont {Fulconis}},
  \bibinfo {author} {\bibfnamefont {J.}~\bibnamefont {Duligall}}, \bibinfo
  {author} {\bibfnamefont {W.}~\bibnamefont {Wadsworth}}, \ and\ \bibinfo
  {author} {\bibfnamefont {P.}~\bibnamefont {Russell}},\ }\href {\doibase
  10.1364/OPEX.13.000534} {\bibfield  {journal} {\bibinfo  {journal} {Opt.
  Express}\ }\textbf {\bibinfo {volume} {13}},\ \bibinfo {pages} {534}
  (\bibinfo {year} {2005})}\BibitemShut {NoStop}%
\bibitem [{\citenamefont {Baker}\ and\ \citenamefont
  {Rochette}(2012)}]{Rochette-fabrication}%
  \BibitemOpen
  \bibfield  {author} {\bibinfo {author} {\bibfnamefont {C.}~\bibnamefont
  {Baker}}\ and\ \bibinfo {author} {\bibfnamefont {M.}~\bibnamefont
  {Rochette}},\ }\href {\doibase 10.1109/JPHOT.2012.2202103} {\bibfield
  {journal} {\bibinfo  {journal} {IEEE Photonics Journal}\ }\textbf {\bibinfo
  {volume} {4}},\ \bibinfo {pages} {960} (\bibinfo {year} {2012})}\BibitemShut
  {NoStop}%
\bibitem [{\citenamefont {Ahmad}\ and\ \citenamefont
  {Rochette}(2012)}]{Rochette-FWM}%
  \BibitemOpen
  \bibfield  {author} {\bibinfo {author} {\bibfnamefont {R.}~\bibnamefont
  {Ahmad}}\ and\ \bibinfo {author} {\bibfnamefont {M.}~\bibnamefont
  {Rochette}},\ }\href@noop {} {\bibfield  {journal} {\bibinfo  {journal} {Opt.
  Express}\ }\textbf {\bibinfo {volume} {20}},\ \bibinfo {pages} {9572}
  (\bibinfo {year} {2012})}\BibitemShut {NoStop}%
\bibitem [{\citenamefont {Lenz}\ \emph {et~al.}(2000)\citenamefont {Lenz},
  \citenamefont {Zimmermann}, \citenamefont {Katsufuji}, \citenamefont {Lines},
  \citenamefont {Hwang}, \citenamefont {Sp\"alter}, \citenamefont {Slusher},
  \citenamefont {Cheong}, \citenamefont {Sanghera},\ and\ \citenamefont
  {Aggarwal}}]{Lenz2000-TPA}%
  \BibitemOpen
  \bibfield  {author} {\bibinfo {author} {\bibfnamefont {G.}~\bibnamefont
  {Lenz}}, \bibinfo {author} {\bibfnamefont {J.}~\bibnamefont {Zimmermann}},
  \bibinfo {author} {\bibfnamefont {T.}~\bibnamefont {Katsufuji}}, \bibinfo
  {author} {\bibfnamefont {M.~E.}\ \bibnamefont {Lines}}, \bibinfo {author}
  {\bibfnamefont {H.~Y.}\ \bibnamefont {Hwang}}, \bibinfo {author}
  {\bibfnamefont {S.}~\bibnamefont {Sp\"alter}}, \bibinfo {author}
  {\bibfnamefont {R.~E.}\ \bibnamefont {Slusher}}, \bibinfo {author}
  {\bibfnamefont {S.-W.}\ \bibnamefont {Cheong}}, \bibinfo {author}
  {\bibfnamefont {J.}~\bibnamefont {Sanghera}}, \ and\ \bibinfo {author}
  {\bibfnamefont {I.~D.}\ \bibnamefont {Aggarwal}},\ }\href@noop {} {\bibfield
  {journal} {\bibinfo  {journal} {Opt. Letters}\ }\textbf {\bibinfo {volume}
  {25}},\ \bibinfo {pages} {254} (\bibinfo {year} {2000})}\BibitemShut
  {NoStop}%
\bibitem [{\citenamefont {Clark}\ \emph {et~al.}(2012)\citenamefont {Clark},
  \citenamefont {Collins}, \citenamefont {Judge}, \citenamefont {M\"{a}gi},
  \citenamefont {Xiong},\ and\ \citenamefont {Eggleton}}]{Clark:12}%
  \BibitemOpen
  \bibfield  {author} {\bibinfo {author} {\bibfnamefont {A.~S.}\ \bibnamefont
  {Clark}}, \bibinfo {author} {\bibfnamefont {M.~J.}\ \bibnamefont {Collins}},
  \bibinfo {author} {\bibfnamefont {A.~C.}\ \bibnamefont {Judge}}, \bibinfo
  {author} {\bibfnamefont {E.~C.}\ \bibnamefont {M\"{a}gi}}, \bibinfo {author}
  {\bibfnamefont {C.}~\bibnamefont {Xiong}}, \ and\ \bibinfo {author}
  {\bibfnamefont {B.~J.}\ \bibnamefont {Eggleton}},\ }\href {\doibase
  10.1364/OE.20.016807} {\bibfield  {journal} {\bibinfo  {journal} {Opt.
  Express}\ }\textbf {\bibinfo {volume} {20}},\ \bibinfo {pages} {16807}
  (\bibinfo {year} {2012})}\BibitemShut {NoStop}%
  \bibitem [{\citenamefont {Chen}(1989)}]{Chen1989}%
  \BibitemOpen
  \bibfield  {author} {\bibinfo {author} {\bibfnamefont {Y.}~\bibnamefont
  {Chen}},\ }\href {\doibase 10.1364/JOSAB.6.001986} {\bibfield
  {journal} {\bibinfo  {journal} {Journal of the Optical Society of America B}\ }\textbf {\bibinfo
  {volume} {6}},\ \bibinfo {pages} {1986} (\bibinfo {year}
  {1989})}\BibitemShut {NoStop}%
\end{thebibliography}
\end{document}